\documentclass[runningheads]{llncs}
\usepackage{graphicx}
\usepackage{wrapfig}
\usepackage{comment}
\usepackage{soul}

\newcommand{\fignum}[1]{\ref{#1}}
\newcommand{\sectnum}[1]{\ref{#1}}
\newcommand{\fig}[1] {Fig.~\fignum{#1}}
 
\newcommand{\sect}[1]{Sec.~\sectnum{#1}}
\newcommand{\eqn}[1]{Eqn.~\ref{#1}}

\usepackage[mathscr]{euscript}

\usepackage[hyphens,spaces,obeyspaces]{url}

\begin{document}
\title{Freeform Islamic Geometric Patterns}
%
%
\author{Rebecca Lin\inst{1,2} \and
Craig S. Kaplan\inst{3}}
\authorrunning{R. Lin and C.S. Kaplan}
%
\institute{Massachusetts Institute of Technology, Cambridge, MA, USA \and
University of British Columbia, Vancouver, BC, Canada \and
University of Waterloo, Waterloo, ON, Canada}
\maketitle              
\begin{abstract}
Islamic geometric patterns are a rich and venerable ornamental tradition. Many classic designs feature periodic arrangements of rosettes: star shapes surrounded by rings of hexagonal petals. We present a new technique for generating `freeform' compositions of rosettes: finite designs that freely mix rosettes of unusual sizes while retaining the aesthetics of traditional patterns. We use a circle packing as a scaffolding for developing a patch of polygons and fill each polygon with a motif based on established constructions from Islamic art. 

\keywords{Islamic geometric patterns \and modular design \and circle packing}
\end{abstract}
\section{Introduction}
\label{sec:introduction}

\begin{figure*}[ht]
\centering
  \includegraphics[width=\linewidth]{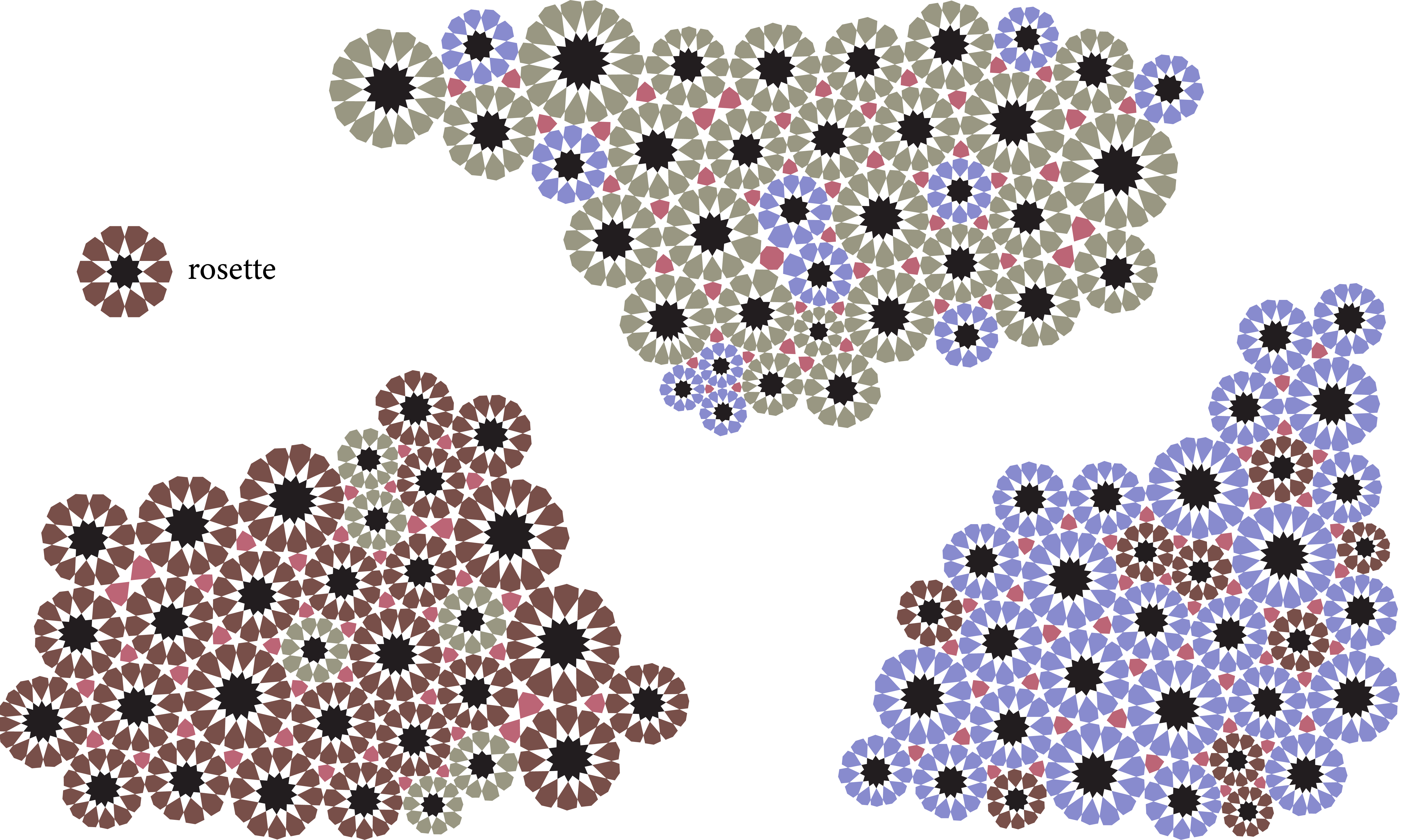}
  \caption{\label{fig:teaser}Freeform designs composed of rosettes of many different orders.}
\end{figure*}

Islamic geometric patterns are a rich and venerable ornamental 
tradition~\cite{Broug2013}. The most iconic of these patterns involve 
periodic arrangements of star shapes, with gaps between them filled by
additional polygons with less symmetry (\fig{fig:hankins-method}). Often, the concave corners of stars are filled with rings of petal-shaped hexagons, yielding composite shapes called \emph{rosettes}~\cite{Lee1987} (\fig{fig:teaser}).

We refer to the number of arms in a star or rosette as its
\emph{order}. These motifs appear in patterns in several
standard orders: multiples of three and four are the most
common due to their compatibility with rotations in periodic
symmetry groups. It is challenging to create patterns that incorporate unusual orders or unusual combinations of orders. For example, considerable geometric sleight-of-hand is required for orders such as~11 and~13 (\fig{fig:hankins-method}b), which are incompatible with crystallographic symmetries~\cite[Pg. 484]{Bonner2017}.

Whether designers employ stars and rosettes of standard or unusual orders, they typically construct \emph{periodic} compositions repeating in two directions in the plane. Repetition is one of the hallmarks of ornamentation: surface decoration on walls and floors, clothing, and objects should be appealing but not distracting. When presented with a periodic pattern, we visually `factor' for a non-repeating kernel and a rule for filling the plane with copies of that kernel. Thus the eye may casually appreciate the pattern without being overwhelmed by it. According to Gombrich~\cite{Gombrich1980}, `aesthetic delight lies somewhere between boredom and confusion', a sentiment echoed by many others~\cite{Hutcheson1726,Arnheim1972,Cromwell2021a}. In decorative contexts, a measure of boredom helps a pattern recede from conscious attention. 

On the other hand, art benefits from a larger dose of confusion. An artwork like a painting is a finite composition that rewards careful study, and so every part of that composition can bear some measure of novelty. In contrast, an infinite Islamic pattern that pleases the eye when elaborated over a wall might lose its appeal if cropped, framed, and hung on that same wall. As an artwork, it would have no natural boundary—no broad composition to guide the eye.

This article presents a technique for constructing `freeform' Islamic
geometric patterns: finite, non-repetitive arrangements of rosettes intended as self-contained compositions rather than as ornamental textures. A few sample compositions appear in \fig{fig:teaser}. Our freeform designs give us significant flexibility to mix and match unusual rosette orders. We move along Gombrich's continuum, away from the boredom of ornamentation and towards the confusion of art. The resulting visual experiment allows us to reimagine the canonical motifs of Islamic geometric patterns in a highly non-traditional setting.

We construct motifs based on an initial polygonal patch using a hybrid of standard techniques (\sect{sec:preliminaries}). We define the overall arrangement of rosettes from a circle packing derived from a triangulation (\sect{sec:freeform-designs}). We show how any circle packing can be converted into a patch of connected polygons (\sect{sec:freeform-patch}). We then inscribe a motif in each polygon and join the motifs together to form the final pattern (\sect{sec:star-motifs}). This technique is robust over a wide range of rosette orders. The designer can control the final pattern by starting with a triangulation of their choosing. We 
support a few additional special effects via `gadgets' that perform
local surgery on the computed circle packing (\sect{sec:gadgets}).
We also adapt our technique to periodic patterns via toroidal circle
packings (\sect{sec:periodicity}).

\section{Related Work} 
\label{sec:related-work}

Artists and mathematicians use many strategies to disrupt the potential monoto\-ny of ornamental Islamic patterns. For example, an artist often introduces mild variations in colours, decorative fills, or calligraphic inscriptions in periodic patterns of otherwise identical stars or rosettes. They also sometimes alter the geometry at the centres of selected rosettes while maintaining outward compatibility with the rest of the pattern.

As the practice of Islamic geometric patterns grew in sophistication,
artists sought to incorporate stars or rosettes of unusual orders into their 
work. The Topkap{\i} scroll, a 15th-century visual guide to the drawing
of Islamic ornament, included a number of patterns with unusual
combinations of stars. Cromwell~\cite{Cromwell2010} analyzed these patterns and articulated rules for their construction. Later, he presented a robust method for assembling patterns from irregular stars with different numbers of points~\cite{Cromwell2013}. That work demonstrated the \textit{wheel construction} (\sect{sec:preliminaries}), which we will detail and use in our method. More recently, Gailiunas~\cite{Gailiunas2020} studied the amount of geometric error that accumulates when juxtaposing otherwise incompatible stars. 

\begin{figure*}[h]
\centering
  \includegraphics[width=\linewidth]{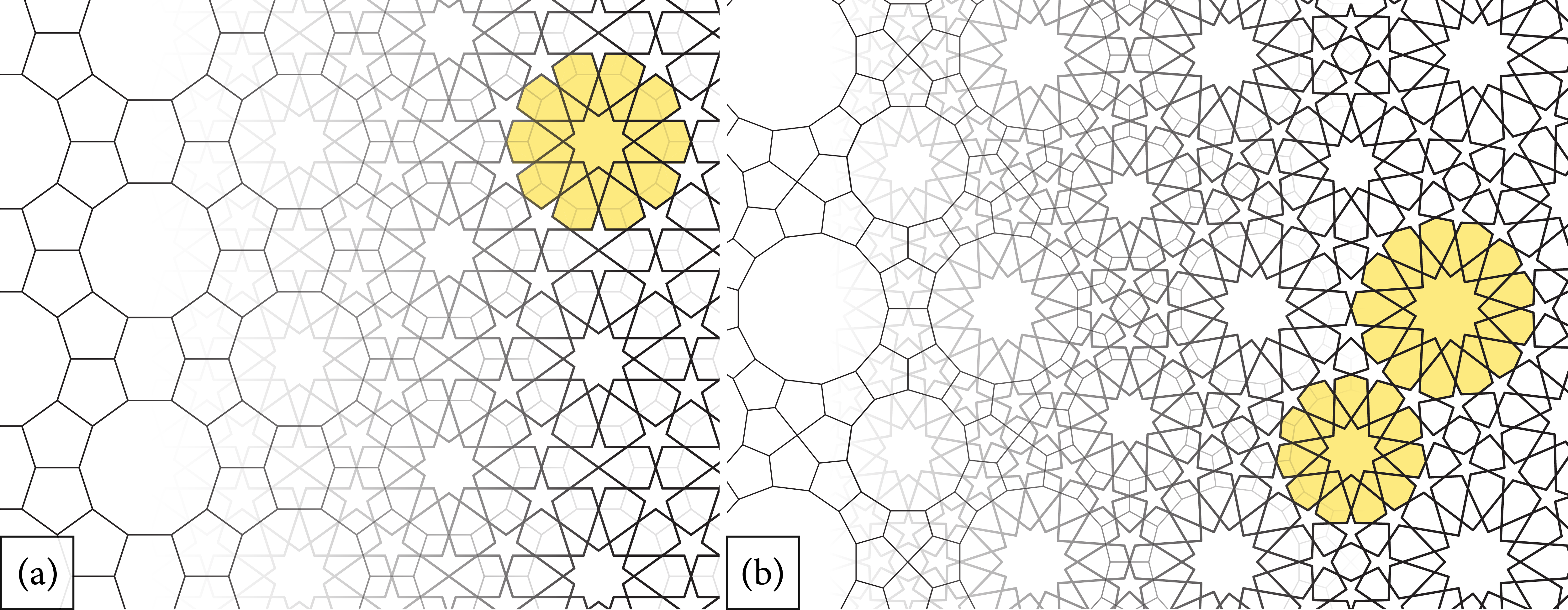}
  \caption{\label{fig:hankins-method}Illustration of polygons-in-contact, with examples of rosettes highlighted. (a) A classical Islamic geometric pattern derived from a tiling by regular decagons, pentagons, and barrel-shaped hexagons. (b) A non-systematic pattern with an underlying tiling by regular 11-gons, 13-gons, and irregular pentagons.}
\end{figure*}

Bonner~\cite{Bonner2017} presented a comprehensive treatment of the modular construction of Islamic patterns. His \textit{polygonal technique}, also known as \textit{polygons-in-contact} (PIC) after
Hankin~\cite{Hankin1925}, builds a motif in every tile of a polygonal
tiling (\fig{fig:hankins-method}). Bonner's book includes a vast collection of patterns with different combinations of stars, including some `non-systematic' patterns that feature stars or rosettes with unusual orders, such as 7, 9, 11, 13, and 14.

Another means of achieving irregularity is to move away from the
Euclidean plane. In Islamic architecture, domes are often
decorated with specialized geometric patterns adjusted to
varying curvature~\cite{Broug2013}. Kaplan and Salesin~\cite{KaplanSalesin2004} demonstrated adapting PIC to produce patterns on the sphere and
in the hyperbolic plane. While repetitive in the mathematical sense,
hyperbolic patterns are necessarily distorted when projected into the Euclidean plane. Kaplan~\cite{Kaplan2009} later presented a more general method for mapping planar patterns with sufficient symmetry, including many Islamic patterns, onto arbitrary surfaces in 3D.

A Moroccan \textit{zellij} design typically features a large central 
star surrounded by radially symmetric constellations of smaller modules~\cite{Castera1999}. These modules are formed from a standard set of individual tile shapes derived from an 8-pointed star. The result is a monumental work containing substantial visual novelty and
appeal. The puzzle of creating such designs is more combinatorial
than geometric: the artist seeks new discrete configurations of a
fixed set of shapes. Recently, Kaplan~\cite{Kaplan2022} presented an algorithm for the procedural generation of small zellij compositions,
which shares some aesthetic goals with our work.

Modern mathematics allows us to produce patterns that are orderly
without being periodic. Many techniques have been proposed that use
substitution tilings or quasiperiodicity to guide the placement of
Islamic motifs~\cite{BP2012a,BP2012b,Castera2010,Castera2016,WR2007}.  Some
researchers have even credited ancient designers with an explicit
understanding of quasiperiodicity~\cite{AlAjlouni2012,LuSteinhardt2007},
though such claims are controversial~\cite{Cromwell2015}. 
Non-periodic patterns with long-range organization occupy an aesthetic
sweet spot: they advertise global structure, but the precise
nature of that structure is not trivially unravelled by the eye.

In the broader world of computer graphics, researchers have explored
some interactive and automated techniques for laying out small motifs
to create ornamental patterns~\cite{Gieseke2021}. Practical numerical 
algorithms for constructing circle packings are relatively new~\cite{CollinsStephenson2003},
so circle packings have not received much attention as an organizing
tool for pattern design. A notable exception is the work of Hamekasi and
Samavati~\cite{Hamekasi2012}, who use circle packings to guide the placement of motifs in Persian floral designs. Most recently, Brewer et al.\ 
derived circle packings from $k$-uniform tilings and used them as a 
framework in which to inscribe Islamic motifs~\cite{Brewer2022}.
Their technique overlaps somewhat with ours, though they are restricted
to arrangements that can arise naturally from the vertex types and
polygon orders of
the tilings they use as a starting point.

\section{Modular Motif Construction} 
\label{sec:preliminaries}

Many standard techniques for constructing Islamic patterns are
\textit{modular}: they decompose the canvas into disjoint regions
such as disks or polygons and define a procedure for filling every
region with a motif. 
This section summarizes two motif construction techniques that will form the basis of our method.

\begin{figure*}[h]
\centering
  \includegraphics[width=\linewidth]{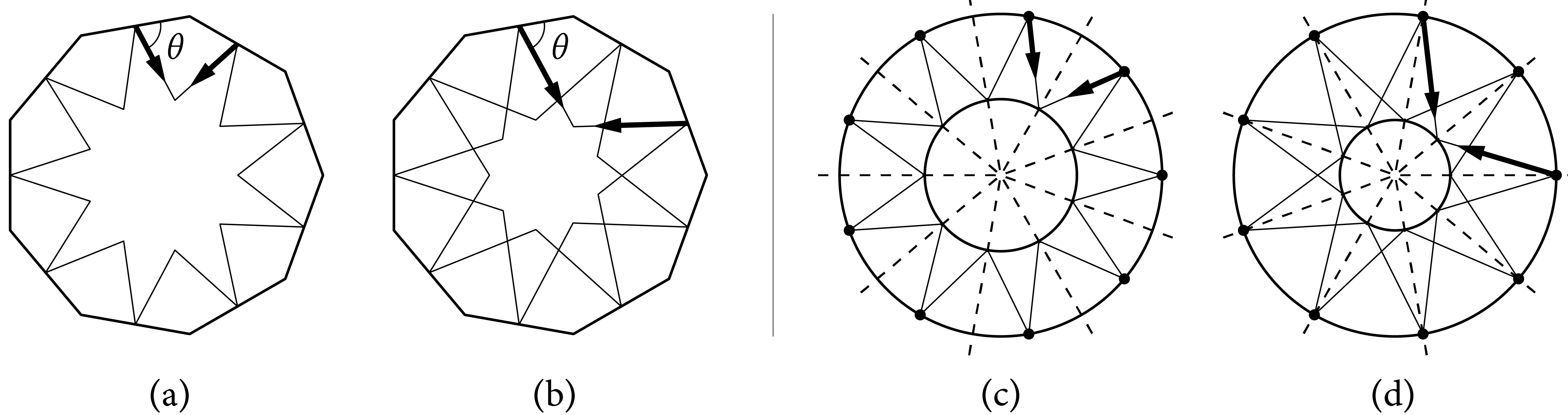}
  \caption{\label{fig:star-construction}Regular 9-pointed stars constructed using PIC (left) and the wheel construction (right). Using PIC, we truncate rays at their first (a) or second (b) intersections with a contact angle of $2\pi/5$. Using the wheel construction, we draw zig-zag paths connecting consecutive points (c) or every other point (d) on the outer circle.}
\end{figure*}

In the \textit{polygons-in-contact} technique (PIC),
the canvas is subdivided into polygons that meet edge-to-edge.
We choose a \textit{contact angle} $\theta\in (0,\pi/2)$.  For every edge of a polygon $P$ in the subdivision, we construct the two rays that grow from the edge's midpoint towards the interior of $P$, rotated by $\pm\theta$ relative to the edge. A motif is formed by truncating these rays where they meet rays from other edges. In simple cases, we need only compute intersections with rays from neighbouring edges 
(\fig{fig:star-construction}a), or from two edges away
(\fig{fig:star-construction}b). A more robust construction requires
heuristics to decide how to truncate rays, such as 
minimizing the total length of the motif's line 
segments~\cite[Sec. 4.4.2]{Bonner2017}.
\fig{fig:hankins-method} shows two patterns created by
constructing motifs for every polygon in a subdivision.

In the \textit{wheel} construction, the modules are circles, each tangent to neighbouring circles in a larger pattern. The construction inscribes a star in every circle. Given a circle $C$ of radius $r$, we first identify a set of points $S$ on its boundary, including the points where $C$ meets its neighbours. We also choose a smaller circle $C'$ with radius $\alpha r$ for a given $\alpha\in(0,1)$, lying in the interior of $C$. Let $p$ and $q$ be two points in $S$. We construct the perpendicular bisector of chord $\overline{pq}$ and find the intersection of that bisector with $C'$. Then we draw line segments from $p$ and $q$ to the intersection. \fig{fig:star-construction} shows stars that emerge when this process is repeated for all pairs of $p$ and $q$ in $S$ that are consecutive (c) or non-consecutive (d). Here we can control the sharpness of the star by varying the scaling ratio $\alpha$ between the radii of the outer and inner circles.

Both of these constructions can produce symmetric $n$-pointed stars.
For PIC, a symmetric star is produced when 
$P$ is regular; for the wheel construction,
we require the points in $S$ to be distributed evenly around $C$, and for
$C$ and $C'$ to be concentric. We can convert between PIC's $\theta$
and the wheel construction's $\alpha$ in this case. Stars (a) and (c) in \fig{fig:star-construction}
are related by

\begin{equation}
    \alpha = 1 - \frac{\sin{\left(\pi/n\right)}\sin{\theta}}{\sin{\left(\pi(n+2)/2n - \theta \right)}}
	\label{eqn:wheel-first}
\end{equation}
and stars (b) and (d) are related by
\begin{equation}
    \alpha = 1 - \frac{2\sin{\left(\pi/n\right)}
        \sin{\left(\pi(n-2)/2n\right)}
        \sin{(\theta - \pi/n)}}
        {\sin{\left(\pi/2 + 2\pi/n - \theta \right)}}.
    \label{eqn:wheel-radius}
\end{equation}

Empirically, the wheel construction works well when forming a star whose
points lie on a
common circle: it degrades gracefully as the point distribution becomes
uneven.
PIC is a better choice for small polygons whose irregularity is harder to
characterize. We shall use both in our method. Note that neither of these techniques explicitly constructs rosettes.
Although explicit rosette constructions exist~\cite{Lee1987}, we will allow
rosettes to emerge as a by-product of the polygonal decompositions we use
as a basis for motif construction, as in the examples of
\fig{fig:hankins-method}.

\section{Freeform Designs} 
\label{sec:freeform-designs}

\begin{figure*}[h!]
\label{fig:overview}
\centering
  \includegraphics[width=\linewidth]{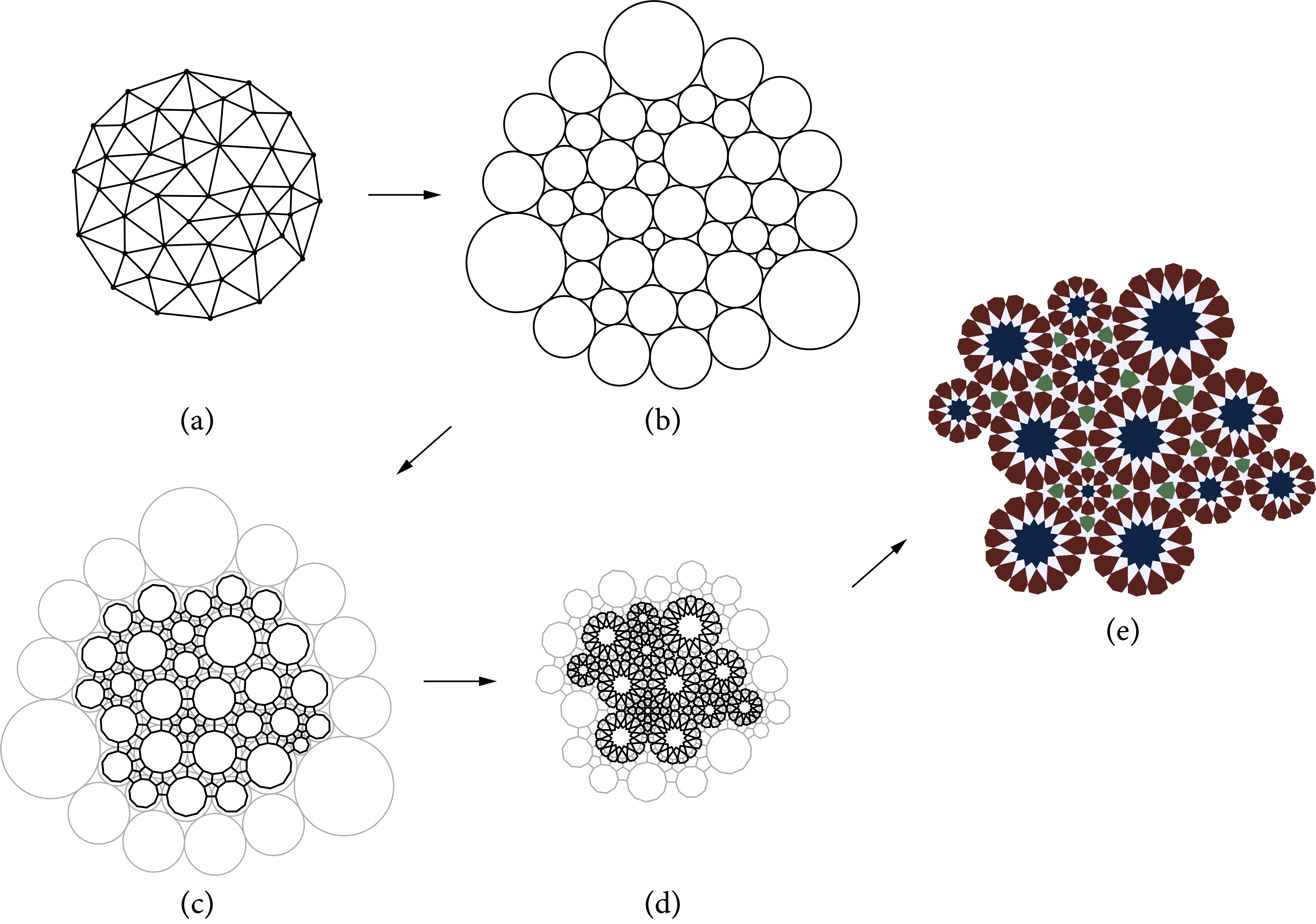}
  \caption{Our method takes a complex (a) and computes a circle packing (b). Then it forms a freeform patch of polygons (c), from which it develops motifs that form a seamless constellation (d) that may be styled (e).}
\end{figure*}

In this section, we present the steps that make up our main technique
for constructing finite, freeform compositions of rosettes. The steps
are visualized in \fig{fig:overview}. We begin with an arbitrary
\textit{simplicial complex}~(a), which induces a \textit{circle packing}~(b). 
Based on the circle packing, we construct a polygonal \textit{patch}~(c),
comprising large cyclic polygons separated by smaller irregular pentagons.
Finally, we use a combination of PIC and the wheel construction to define
\textit{motifs} for each polygon~(d), and optionally render
the design~(e). In the following subsections, we 
describe each of these steps in detail.

\subsection{Complex}

The main input to our technique is a \textit{complex}, more formally
a planar, simply connected, pure simplicial
2-complex $\mathcal{K}$.  In simpler terms, we may regard
$\mathcal{K}$ as a collection of non-overlapping triangles in the plane,
meeting edge-to-edge. The union of the triangles defines a region
the plane---a simple polygon.
We will refer to
the vertices, edges, and faces of the complex. We distinguish between its
\textit{boundary vertices}, which lie on the simple polygon, 
and \textit{interior vertices}, which lie interior to the polygon.

We may construct input complexes in numerous ways. It is easy to
author them manually by placing and connecting vertices. They can
also be generated procedurally, such as by computing the Delaunay 
triangulation of a point set. 

\subsection{Circle Packing}

Let $\mathcal{K}$ be a complex with $n$ vertices.
A \textit{circle packing}
for $\mathcal{K}$ is a collection of non-overlapping circles
$\{C_1,\ldots,C_n\}$ whose tangencies echo the combinatorial structure
of $\mathcal{K}$.  Each circle $C_i$ corresponds with vertex
$v_i$ of the complex, and two circles $C_i$ and $C_j$ are externally
tangent if and only if $v_i$ and $v_j$ are connected by an edge in 
$\mathcal{K}$.
The Discrete Uniformization Theorem guarantees that a circle packing
exists for any given complex $\mathcal{K}$~\cite{Stephenson2005}.
Although the circle packing's connectivity will be identical to
that of its complex, they will generally not be equivalent 
\textit{geometrically}: the locations and sizes of the circles
are not directly related to the locations of the vertices in the complex, or
to the shapes of its triangles.

Collins and Stephenson~\cite{CollinsStephenson2003} 
describe a simple numerical algorithm that
computes circle packings through
iterative adjustments of an initial assignment of radii to the
$C_i$.  The radii of boundary circles must be further constrained
with additional boundary conditions. The simple Python script by
Eppstein~\cite{Eppstein} 
accepts explicit values for boundary radii.
Given a boundary vertex of degree $n$, our implementation chooses
a radius $r$ for a circle that would be perfectly surrounded by
$2n-2$ unit circles, giving 
$r=(1-\sin\phi)/\sin\phi$, where $\phi=\pi/(2n-2)$.


\subsection{Polygonal Patch}
\label{sec:freeform-patch}

A \textit{patch} is a finite set of polygons with disjoint interiors
whose union is a topological disk. Given a circle packing, we
construct a patch that has a large cyclic polygon (i.e., a polygon
whose vertices lie on a common circle) associated with each circle,
separated from other cyclic polygons by haloes of pentagonal `filler
polygons'. By design, these polygons can serve as scaffolding
for building motifs typical in Islamic geometric patterns.

\begin{figure}[h]
\centering
  \includegraphics[width=\linewidth]{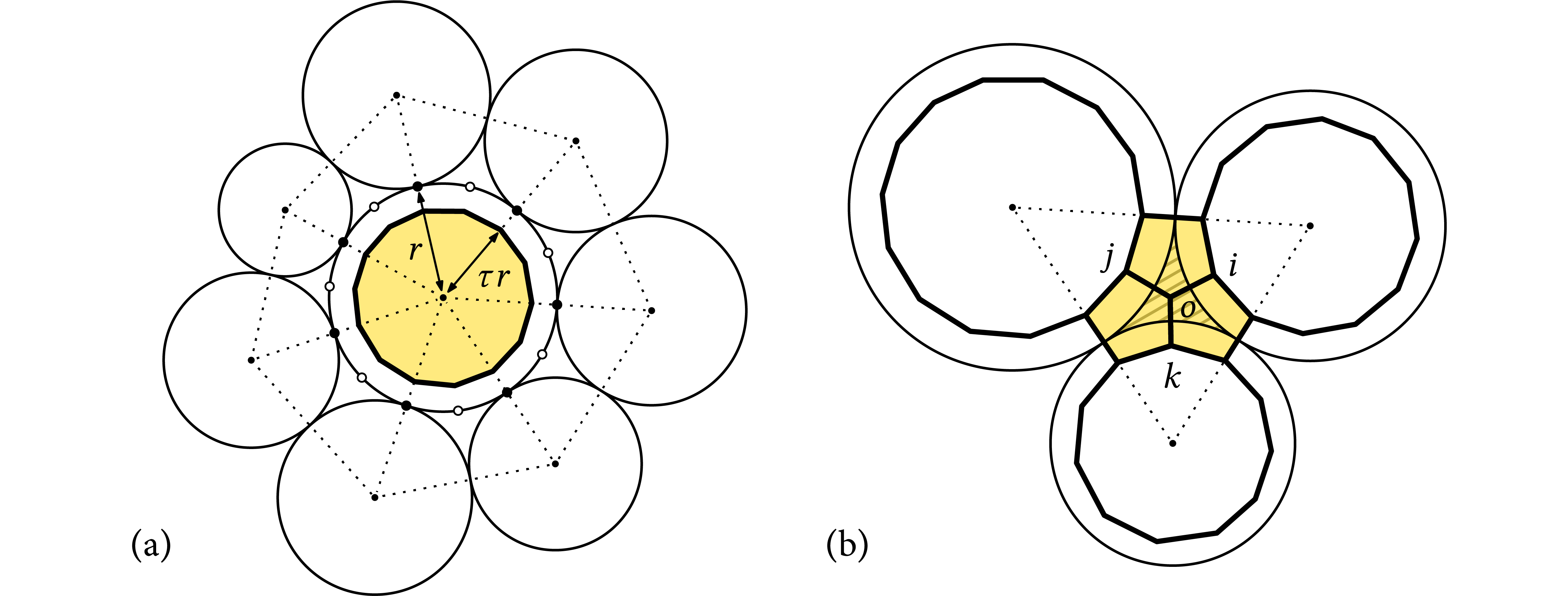}
  \caption{\label{fig:freeform-patch}Constructing a polygonal patch from a circle packing: 
  	we create a cyclic polygon for every circle (a), and fill the gaps
	between three mutually tangent circles with trios of irregular pentagons (b).}
\end{figure}

Let $C$ be an interior circle in a circle packing, and let $k$ be the
degree of the vertex associated with $C$ in the complex. As illustrated
in \fig{fig:freeform-patch}a, we construct
a cyclic $2k$-gon $P$ in the interior of $C$. To begin, we set the vertices
of $P$ to be the $k$ points of tangency between $C$ and its neighbours,
together with the midpoints of the minor arcs of $C$ connecting adjacent
tangency points. Now
let $\tau\in(0,1)$ be a user-selected scaling factor. Scale $P$ relative
to the centre of $C$ by a factor of $\tau$, and add the scaled polygon to
the patch. By default, we use $\tau=0.8$,
a choice that we discuss in \sect{sec:details}.

The gaps between circles in the packing are triangular regions bounded
by arcs of three mutually tangent circles. Let $C_i$, $C_j$, and $C_k$ be one such trio of circles. We divide the space between their cyclic polygons into three new pentagons, as shown in \fig{fig:freeform-patch}b, by drawing edges connecting vertices of cyclic
polygons.
Three outer line segments pass through the pairwise tangencies of the
circles. Three inner segments connect arc midpoints to a new point $o$, 
the incentre of the triangle formed from the centres of $C_i$, $C_j$, and
$C_k$.

\subsection{Motif Construction} 
\label{sec:star-motifs}

\begin{figure}[ht]
\centering
  \includegraphics[width=\linewidth]{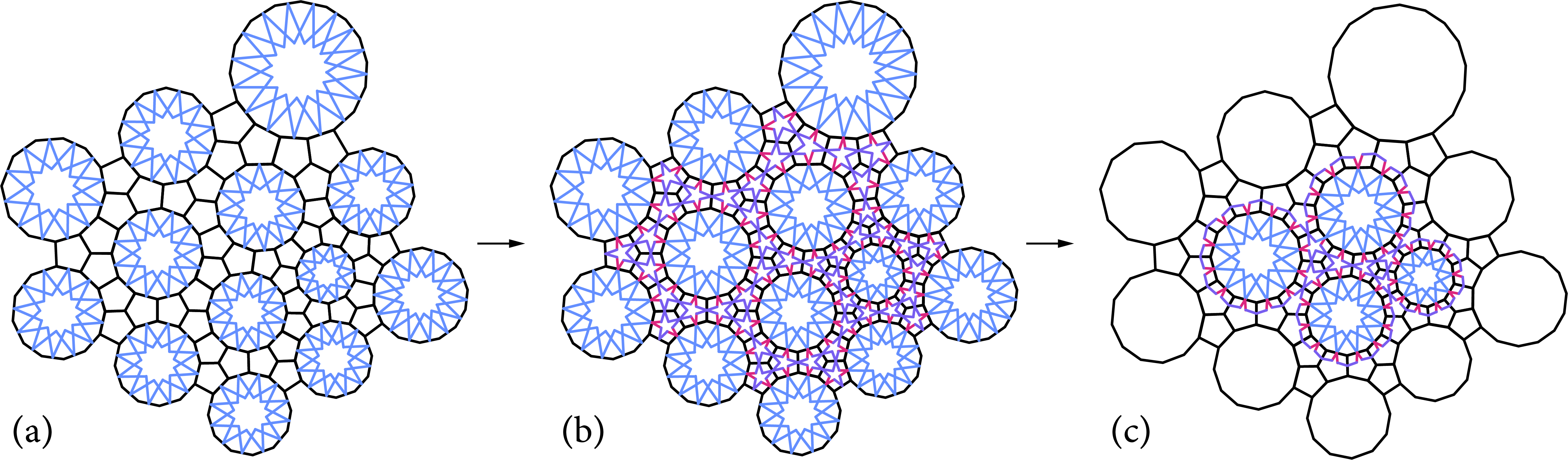}
  \caption{\label{fig:motif-construction}To construct a design from a patch, we use the wheel construction
  to create a star in every cyclic polygon~(a) and apply PIC to build motifs
  for filler polygons~(b).  Optionally, we remove the outer layers of geometry
  to extract an arrangement of whole rosettes~(c).}
\end{figure}

The final step in our process is to construct a motif for every
polygon in the patch produced in the previous step. Here we apply
both the wheel construction and PIC, depending on the type of polygon
being decorated. Our large cyclic polygons yield motifs that garner
attention. We safeguard the quality of these motifs by exploiting
the robustness of the wheel construction in their development
(\fig{fig:motif-construction}a).  We then use PIC for the more
unpredictable filler pentagons (\fig{fig:motif-construction}b).
Optionally, we remove motif segments around the boundary of the resulting composition, paring it down to a core of whole rosettes
(\fig{fig:motif-construction}c).  Our construction depends on a
single global contact angle $\theta$, as described in
\sect{sec:preliminaries}. By default, we use $\theta=2\pi/5$, the
angle for which PIC would inscribe a perfect pentacle
in a regular pentagon.

Let $C$ be an interior circle in the packing with centre $o$ and $k$
points of tangency. Let $P$ be the cyclic $2k$-gon associated with $C$
in the patch. We use the wheel construction to build
a star centred at $o$
whose outer points lie at the edge midpoints of $P$. Generally,
these midpoints do not lie on a common circle, but the wheel construction 
is tolerant of small deviations in their distances from $o$.
Let $r_C$ be the radius of $C$, and define $r$ to be
$r_C\cos(\pi/2k)$.  The value $r$ approximates the radius of an
inscribed circle meeting $P$'s edge midpoints, an approximation that 
converges on the correct value
when $P$ is regular.  Now compute $\alpha$ by plugging the
user-supplied contact angle $\theta$ into \eqn{eqn:wheel-radius}, and let
$C'$ be a circle with center $o$ and radius $\alpha r$.  The radius of
$C'$ is chosen to ensure that the contact angles at the points of the star
approximate $\theta$. We apply the
wheel construction using the edge midpoints of $P$ and the inner circle
$C'$, connecting every other star point as in \fig{fig:star-construction}d.

It remains to build motifs for the filler pentagons. Let $Q$ be one such
pentagon. As in PIC, construct a pair of rays emanating from the midpoint
of every edge of $Q$, and truncate them where they intersect rays growing
from neighbouring edges. If an edge $e$ of $Q$ is adjacent to a cyclic polygon,
then we choose contact angles that yield rays parallel to the star
edges meeting across $e$ (\fig{fig:motif-construction}b, red).
These angles may not be symmetric across the perpendicular bisector of $e$,
but the discrepancy is small in practice.
If $e$ is adjacent to another pentagon, on the other hand, then we use
$\theta$ as the contact angle for its rays 
(\fig{fig:motif-construction}b, purple).

In summary, our method uses a patch to construct a constellation
of localized motifs that combine to form familiar visual elements:
rosettes. By our application of the Discrete Uniformization Theorem,
each rosette corresponds to a vertex in the triangulation
$\mathcal{K}$, and two
rosettes are adjacent if and only if their vertices share an edge
in $\mathcal{K}$. The order of a rosette is twice the degree of its
associated vertex.

\section{Gadgets} 
\label{sec:gadgets}

The basic technique of the previous section can produce a wide
variety of freeform designs with combinations of rosettes of different
orders. However, some configurations found in traditional Islamic
geometric patterns remain out of reach, most obviously because
we define only one way to fill the triangular gaps between cyclic
polygons. In this section, we introduce two \textit{gadgets} that
help us recover some of that variety, increasing the visual 
intrigue of our designs. Gadgets are small subgraphs
with labelled vertices that can be incorporated into a complex.
These vertices then determine local clusters of polygons during
patch construction, overriding the polygons of \sect{sec:freeform-patch}.

\begin{figure}[h]
\centering
  \includegraphics[width=\linewidth]{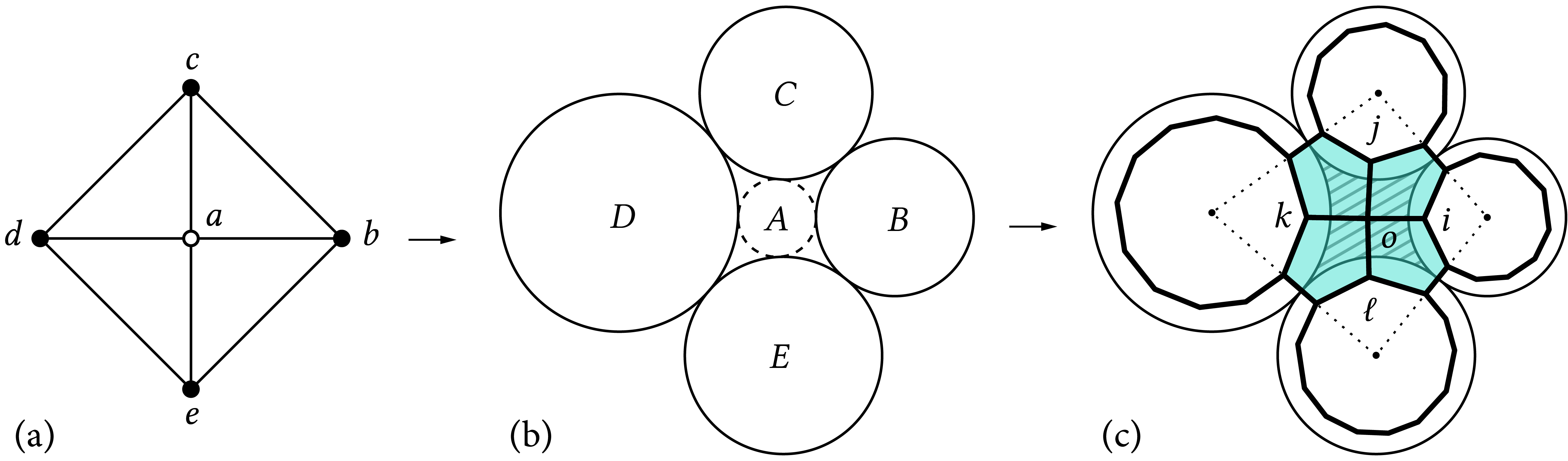}
  \caption{\label{fig:bowtie-gadget}The square gadget~(a) produces a circle packing~(b) from which we derive four filler pentagons~(c).}
  \vspace{5pt}
  \includegraphics[width=\linewidth]{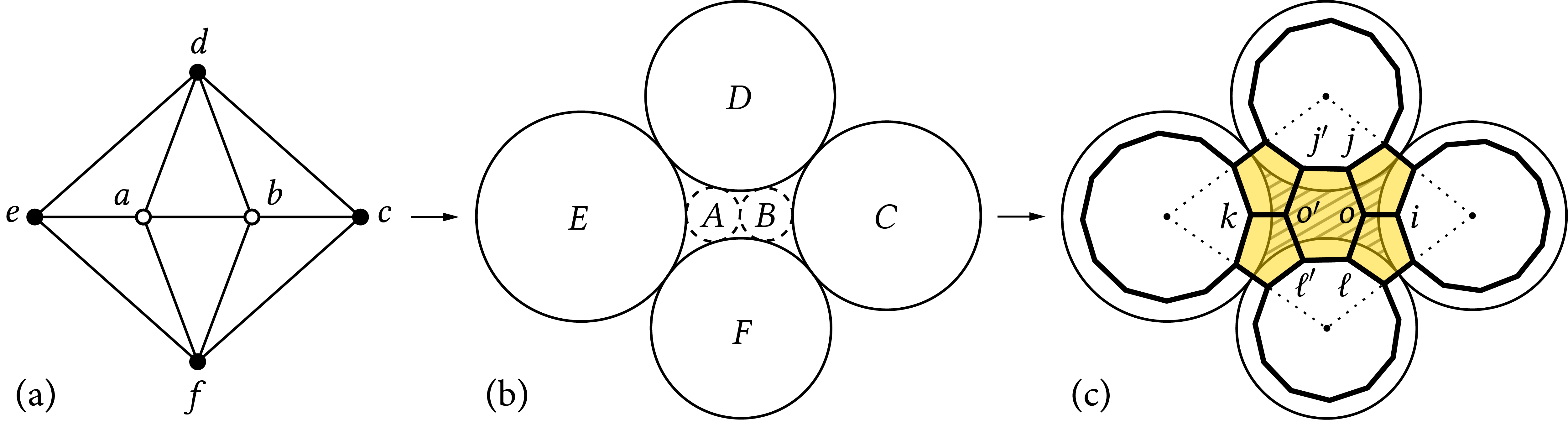}
  \caption{\label{fig:square-gadget}The bowtie gadget~(a) produces a circle packing~(b) from which we derive a cluster~(c) of four filler pentagons and a barrel-shaped hexagon.}
\end{figure}

A \emph{square gadget} is a 5-vertex subgraph with a central vertex
$a$ of degree 4, as shown in \fig{fig:square-gadget}a. 
Given a complex containing a copy of the square gadget, 
we obtain a circle packing containing a cluster of circles like the one
shown in \fig{fig:square-gadget}b, where circle $A$ is associated with
vertex $a$.
When building the patch,
we remove $A$ from the circle packing and tile the hole left behind
with four pentagons, as shown in \fig{fig:square-gadget}c.
The new point $o$ is the mean of vertices $i$ ,$j$, $k$, and $\ell$.
Our motif construction will produce a squarish region surrounded by four rosettes containing a central octagon.

\vspace{-10pt}

A \emph{bowtie gadget} is a 6-vertex subgraph with two central vertices
$a$ and $b$ of degree 4, as shown in \fig{fig:bowtie-gadget}a.
As with the square gadget, we remove the corresponding circles $A$ and
$B$ from the circle packing and fill the void with a new configuration
of tiles. First, when constructing a cyclic polygon for circle
$D$ associated with vertex $d$, we divide the minor
arc between the tangencies with $C$ and $E$ into \textit{three} equal
pieces instead of the usual two, yielding vertices $j$ and $j'$ in 
\fig{fig:bowtie-gadget}c. Similarly, we divide $F$'s arc into three,
which gives us vertices $\ell$ and $\ell'$. We then construct a bowtie-shaped arrangement of four pentagons and one barrel-shaped hexagon, as illustrated in \fig{fig:bowtie-gadget}c, where $o$ is the mean of $i$, $j$, and $\ell$ and $o'$ is the mean of $j'$, $k$, and $\ell'$. 




\setlength{\columnsep}{8pt}
\begin{wrapfigure}{r}{0.26\linewidth}
  \vspace{-32.5pt}
  \begin{center}
    \includegraphics[width=\linewidth]{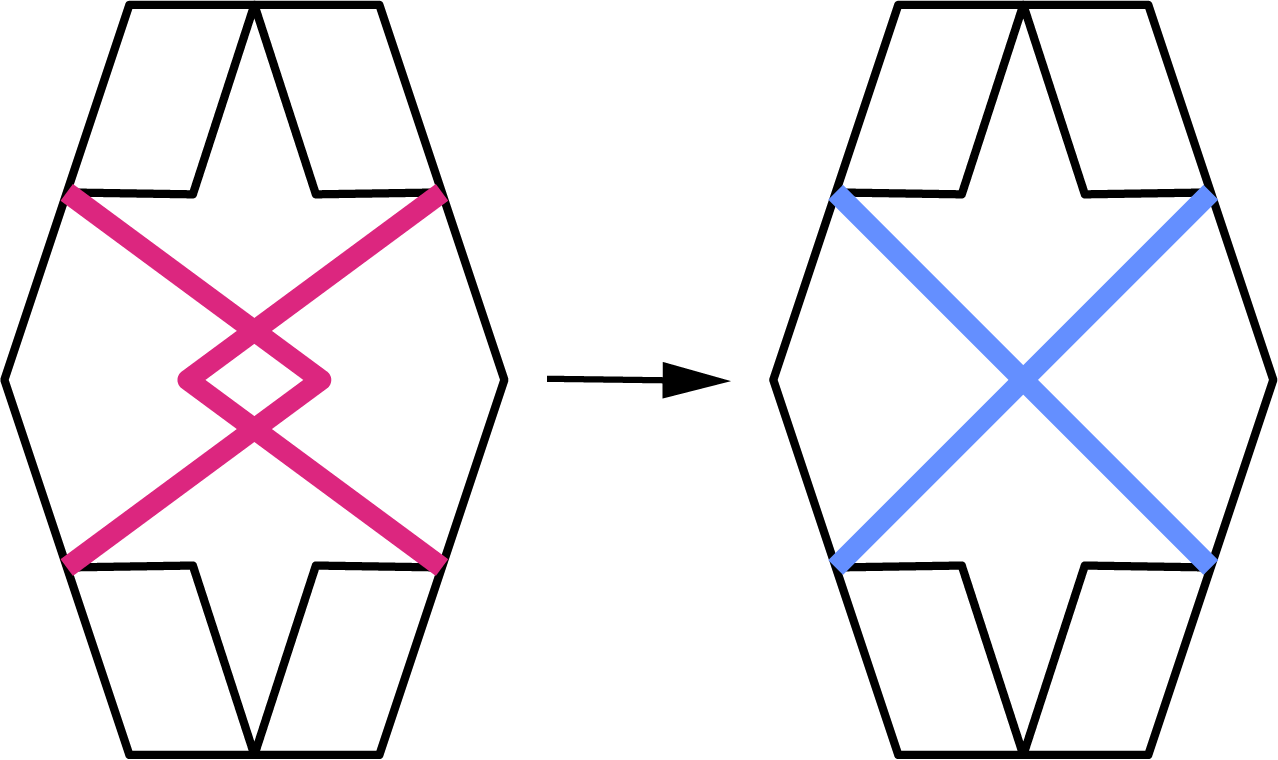}
  \end{center}
  \vspace{-32pt}
\end{wrapfigure} 
When the barrel-shaped hexagon in the centre of the bowtie gadget becomes too
thin, it can produce a motif with a small region of overlap at its centre,
as shown in red in the inset. When these overlaps occur, we replace the red
segments with a perfect `X', shown in blue. The blue segments alter the
contact angles with the edges of the hexagon; we propagate any changes
to the hexagon's four neighbouring pentagons.


\setlength{\columnsep}{8pt}
\begin{wrapfigure}{l}{0.17\linewidth}
  \vspace{-32.5pt}
    \begin{center}
    \includegraphics[width=\linewidth]{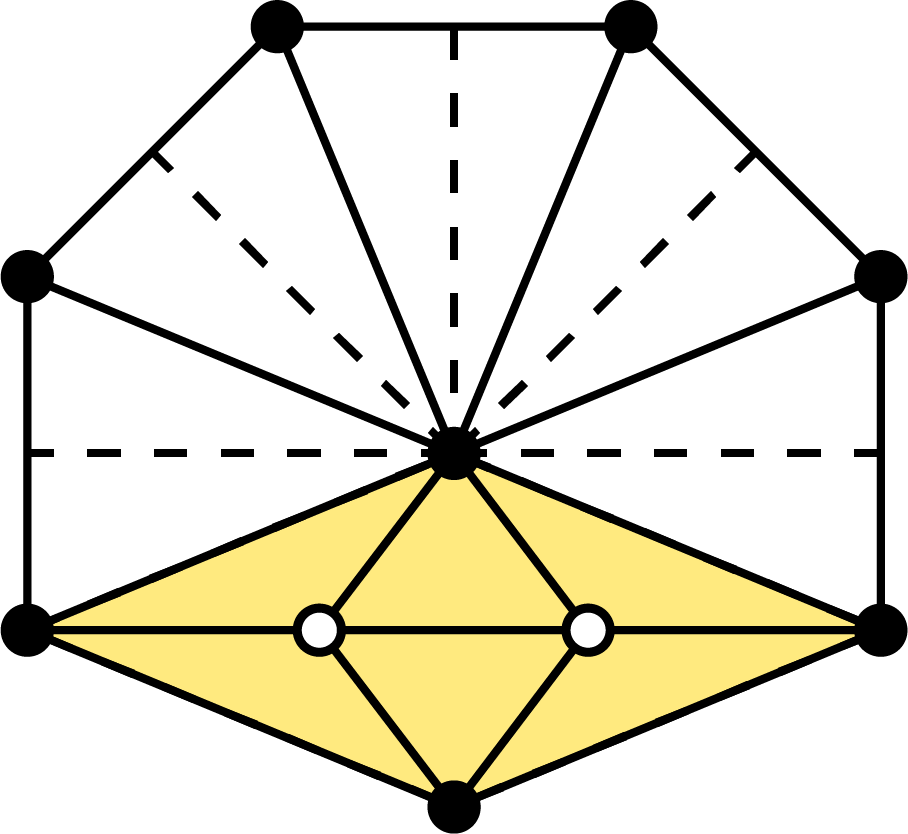}
  \end{center}
    \vspace{-40pt}
\end{wrapfigure} 
Recall that without gadgets, our construction was limited to 
rosettes of even orders. But when a bowtie gadget appears in a complex, 
vertices $d$ and $f$ each contribute three edges
to their corresponding cyclic polygons.  
Therefore, if a complex 
vertex acts as $d$ or $f$ in one such gadget, as in the central vertex
of the subgraph in the inset, that vertex will yield a rosette of
odd order. More generally, we may hang any odd
number of suitably oriented bowtie gadgets from a vertex to obtain an
odd-order rosette. 

\begin{figure}[h!]
\centering
  \includegraphics[width=\linewidth]{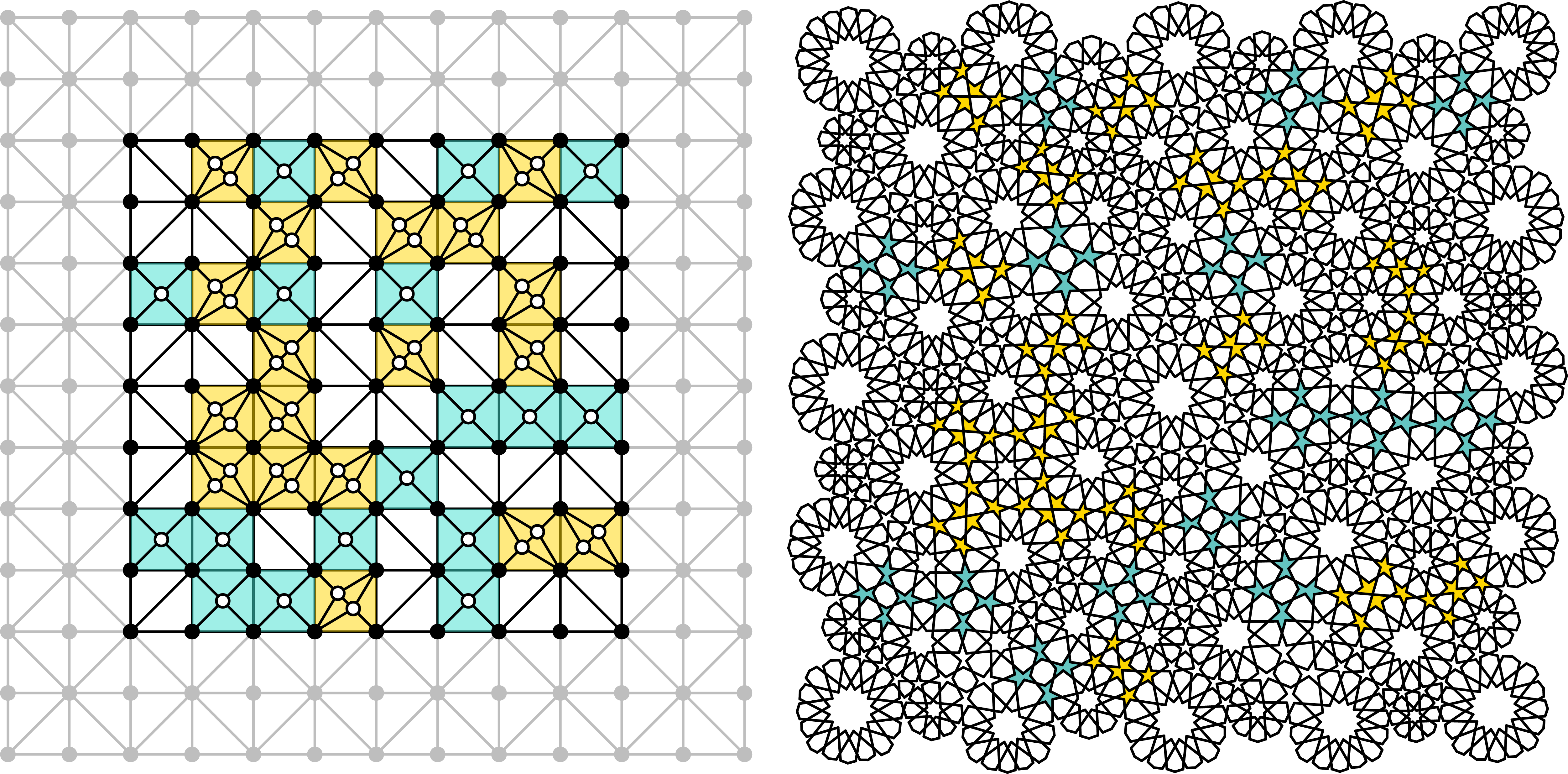}
  \caption{\label{fig:grid-gadgets}A composition based on a square grid, where every square is 
	randomly subdivided with a diagonal, a square gadget, or a 
	bowtie gadget.}
\end{figure}

\fig{fig:grid-gadgets} shows a freeform design constructed from a random arrangement of bowtie and square gadgets. In future work, we hope to explore gadgets beyond these two.

\section{Periodic Patterns} 
\label{sec:periodicity} 

\begin{figure}[h]
\centering
\includegraphics[width=\textwidth]{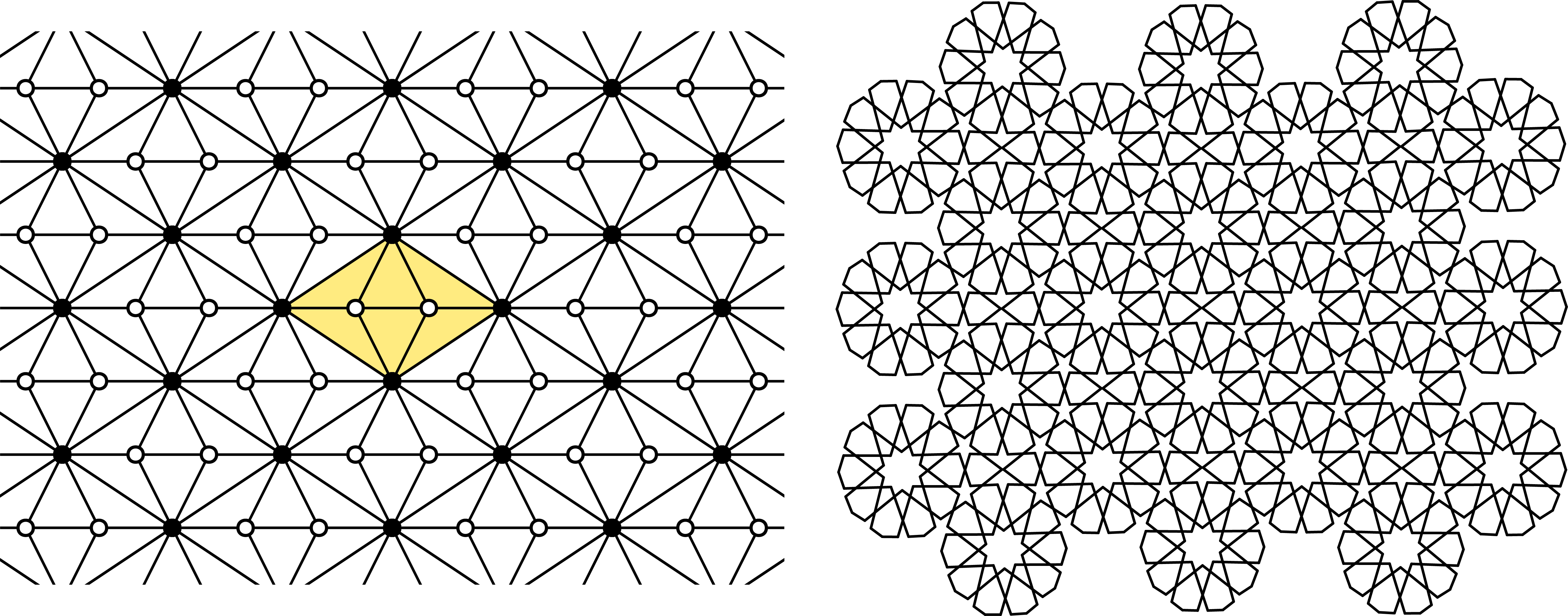}
\caption{\label{fig:freeform-classical}A periodic arrangement of bowtie gadgets (left) 
can be used to generate a freeform version of the
pattern in \fig{fig:hankins-method}a (right).}
\end{figure}

While the focus of our technique is the creation of finite, freeform
compositions, we have also examined its ability to produce more
orderly designs. 
For example, a finite subset of a periodic arrangement of bowtie gadgets (\fig{fig:freeform-classical}, left) yields an approximation of the 
decagonal design in \fig{fig:hankins-method}a (\fig{fig:freeform-classical}, right).  Other periodic
arrangements of triangles and gadgets can reproduce different classic designs.
However, because of flexibility in the circle packing algorithm, these freeform
designs could contain rosettes of continuously varying scales.

We can extend our technique to generate truly periodic patterns
by generalizing the Discrete Uniformization Theorem beyond the
Euclidean plane.  In particular, if $\mathcal{K}$ is embedded on
a torus, then the theorem guarantees the existence of a circle
packing in the torus's intrinsic metric~\cite[Ch. 9]{Stephenson2005}.
The circle packing algorithm is, in some sense, even simpler in this case
because there is no longer any need for explicit boundary conditions:
every circle is completely surrounded.
The torus can then be cut open and unrolled, yielding a finite
collection of circles that can be stamped out to produce
an infinite periodic packing.  

\begin{figure}[ht]
\centering
  \includegraphics[width=\linewidth]{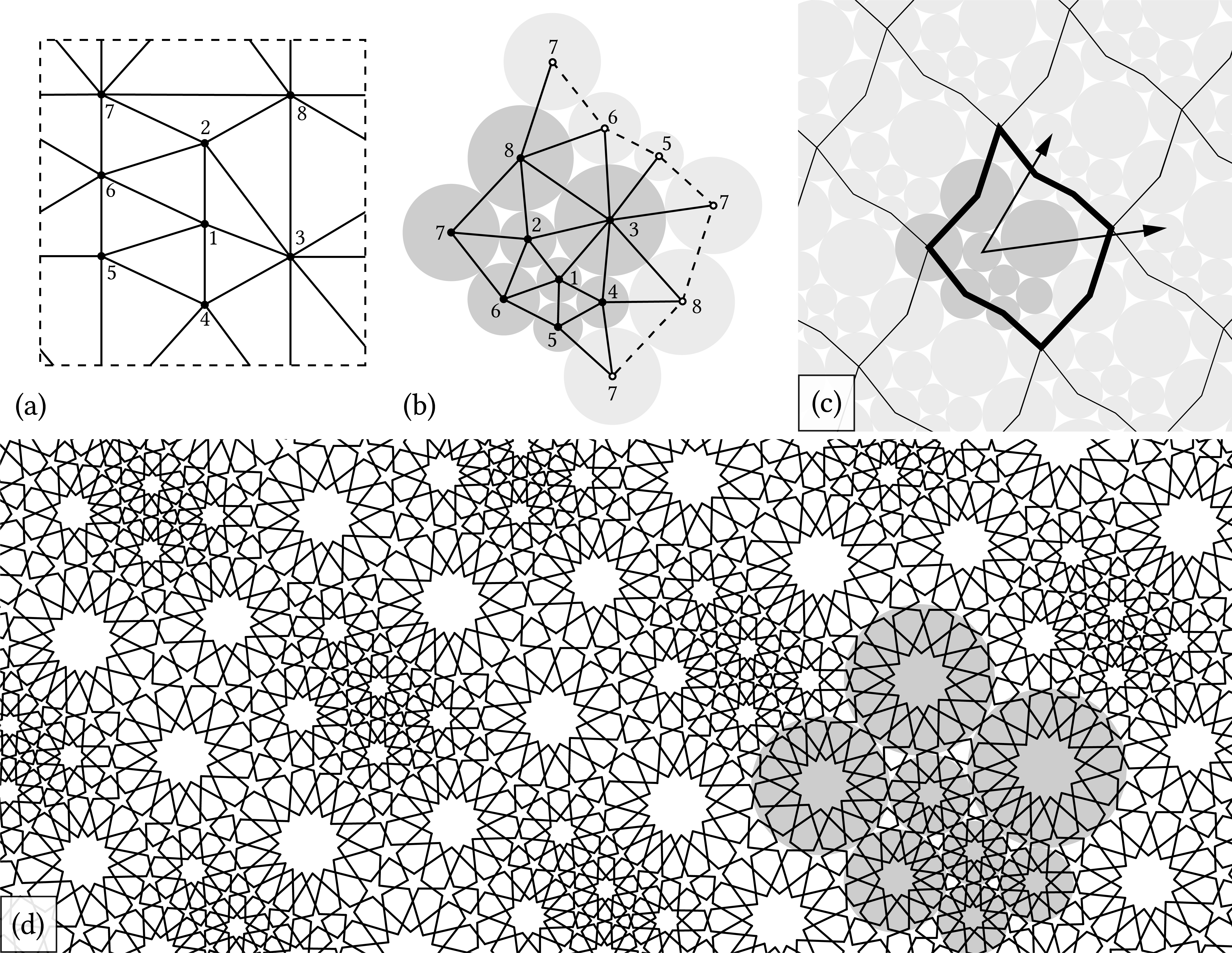}
  \caption{\label{fig:periodic-pattern-complex}A triangulation
  drawn on a square with periodic boundary conditions~(a) is used to generate
  a circle packing~(b) that covers the plane through translation~(c).  We construct
  motifs to obtain a periodic Islamic geometric pattern~(d).}
\end{figure}

\fig{fig:periodic-pattern-complex} gives an example of a periodic pattern 
generated from a simplicial complex embedded on a torus.  
The light grey disks in  \fig{fig:periodic-pattern-complex}b should be interpreted as 
translated copies
of the dark grey disks with the same indices. 
The numerical circle packing algorithm yields
a layout that tiles the plane by translation (\fig{fig:periodic-pattern-complex}c),
from which we can create a periodic pattern with rosettes of orders
10, 12, 14, and 16.

Future work could explore the analogous extensions of this technique
to other spaces, such as the sphere and the Poincar\'e disk model
of the hyperbolic plane~\cite{KaplanSalesin2004}.

\begin{figure}[ht]
\centering
  \includegraphics[width=\linewidth]{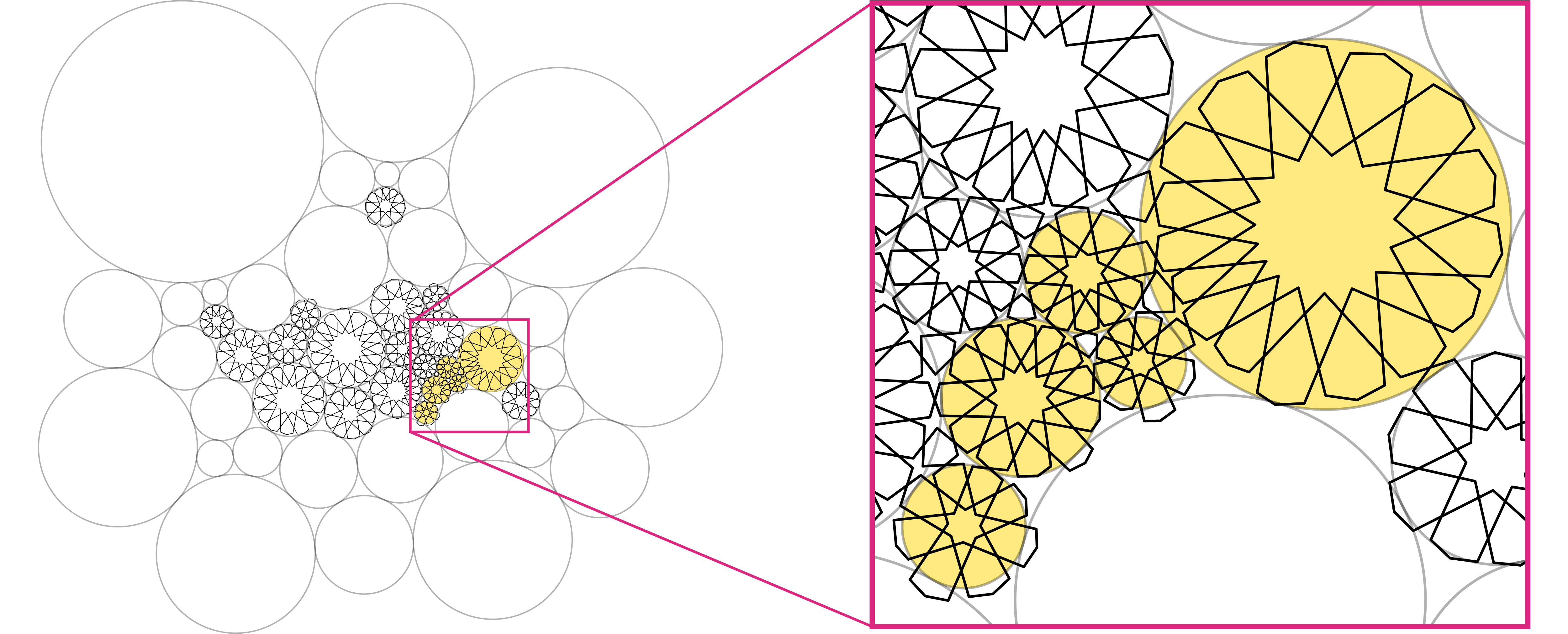}
  \caption{\label{fig:circle-selection}Circles near the boundary of a packing can lead to distorted stars and rosettes
  (right). We discard outer circles, which can sometimes partition
  a design into multiple connected components (left).}
\end{figure}

\section{Discussion}
\label{sec:details} 

In this section, we discuss some of the details of our technique, including alternative approaches that we considered during its development. Some of these alternatives may offer opportunities for future work.

\textbf{Selecting interior circles. }
We typically do not generate a motif for every circle in the packing. Boundary circles, and circles adjacent to them, can differ substantially in size from their neighbours. These variations can propagate through the rest of the construction and produce unacceptably distorted motifs, such as uneven rosette petals (\fig{fig:circle-selection}). Future work could consider ways to optimize the geometry
of the circle packing to serve patch and motif construction. 
For now, we omit outer layers of circles in our final designs. 
Note that this approach may separate the design into multiple connected components, in which case we simply keep the largest component.

Beyond these technicalities, we can be selective for aesthetic reasons. Having the freedom to craft the shape of a design provides opportunities to create interesting compositions (\fig{fig:generative}).

\setlength{\columnsep}{8pt}
\begin{wrapfigure}{r}{0.38\linewidth}
  \vspace{-12pt}
  \begin{center}
    \includegraphics[width=\linewidth]{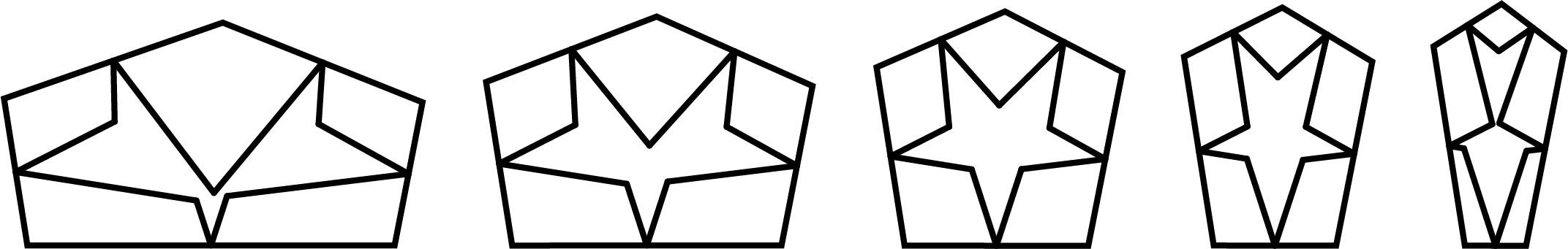}
  \end{center}
  \vspace{-32pt}
\end{wrapfigure} 
\textbf{Choosing a scale factor. } Recall that the parameter~$\tau$ controls the scale of each cyclic polygon relative to its circle, which in turn affects the shapes of filler pentagons. As shown in the inset, the quality of a motif generated within a filler pentagon decreases as that pentagon deviates from regularity. Thus we seek to choose $\tau$ to minimize the total deviation across a design.

\begin{figure}[ht]
\includegraphics[width=\linewidth]{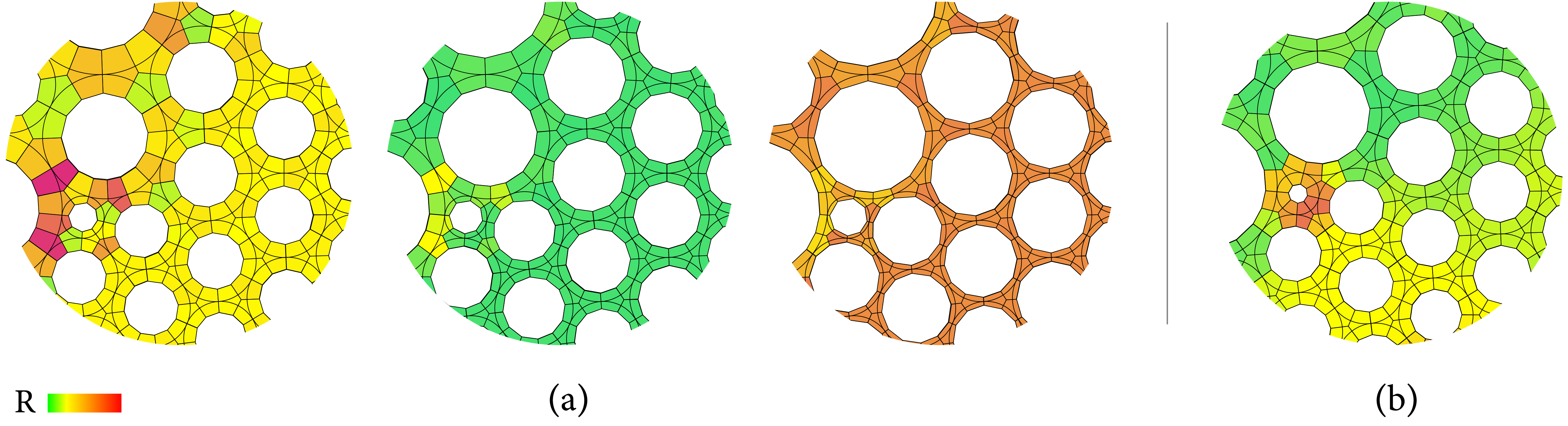}
  \caption{\label{fig:scale-factor}A patch with pentagons coloured by their deviations (red) from regularity (green). In (a), cyclic polygons are scaled by $\tau = 0.7$, $0.8$, and $0.9$, showing that $0.8$ produces good quality overall. In (b), they are offset by a fixed amount, with less consistent results.}
\end{figure}

To gauge the deviation of a pentagon $Q$ from regularity, we adopt a continuous symmetry measure by Zabrodsky et al.~\cite{ZPA1992}, which quantifies the minimal distance that the vertices of $Q$ must travel to form a regular pentagon. Let the error of a 
 freeform patch be the average deviation of its pentagons from regularity. We can compute this error for a range of closely-spaced $\tau$ values and choose the one with minimal error (\fig{fig:scale-factor}a).
 Over a range of circle packings, we
 see significant deviation outside the range $(0.7,0.9)$ and find that $\tau=0.8$ produces satisfactory results, as shown throughout this work.

In the future, we hope to investigate other measurements of polygon quality
in order to produce patches that are closer to ideal. For example, PIC can often
produce a satisfactory motif in a polygon that has lower-order symmetries while
not being fully regular.

As an alternative to treating $\tau$ as a scaling factor, we also considered \textit{offsetting} cyclic polygons by a constant inward distance~$\tau$. However, we found that this approach was not as successful in producing high-quality pentagons (\fig{fig:scale-factor}b). With either interpretation of $\tau$, the quality is the poorest for pentagons adjacent to two cyclic polygons of widely different radii. Hamekasi and Samavati note this issue as well~\cite{Hamekasi2012}, and mitigate it
by avoiding complexes containing neighbouring vertices of widely varying degrees.
In future work, we would like to develop a global optimization that chooses a different scaling factor for every cyclic polygon so as to maximize the overall quality of all filler pentagons.

\setlength{\columnsep}{7pt}
\begin{wrapfigure}{r}{0.39\linewidth}
  \vspace{-32pt}
  \begin{center}
    \includegraphics[width=\linewidth]{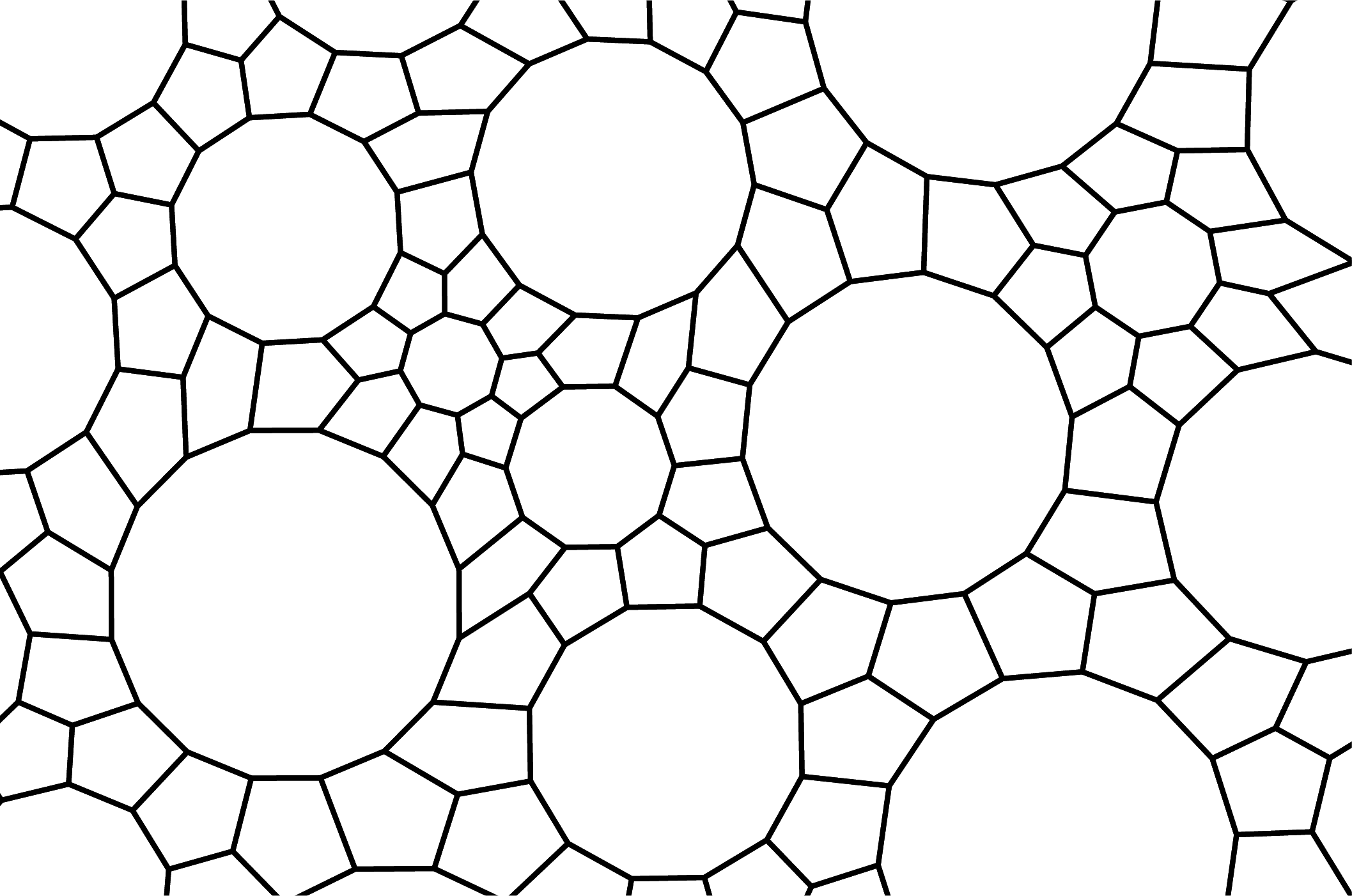}
  \end{center}
  \vspace{-32pt}
\end{wrapfigure} 
\textbf{Cyclic vs. regular polygons. } It is tempting to construct regular polygons in place of cyclic polygons, as these would yield perfectly symmetric stars as motifs. Using the aforementioned regularity measurement~\cite{ZPA1992}, we fit a regular polygon $\hat{P}$ to each cyclic polygon $P$ generated in \sect{sec:freeform-patch}, and centre $\hat{P}$ on the circumcentre of $P$. The result for $\tau=0.8$ is
shown in the inset. This approach prioritizes the quality of large, prominent stars. However, it yields distorted pentagons whose motifs self-intersect. By choosing cyclic polygons rather than regular polygons, our algorithm sacrifices some quality in large stars for the sake of creating feasible connections between them.

\begin{figure}[h!]
\centering
  \includegraphics[width=\linewidth]{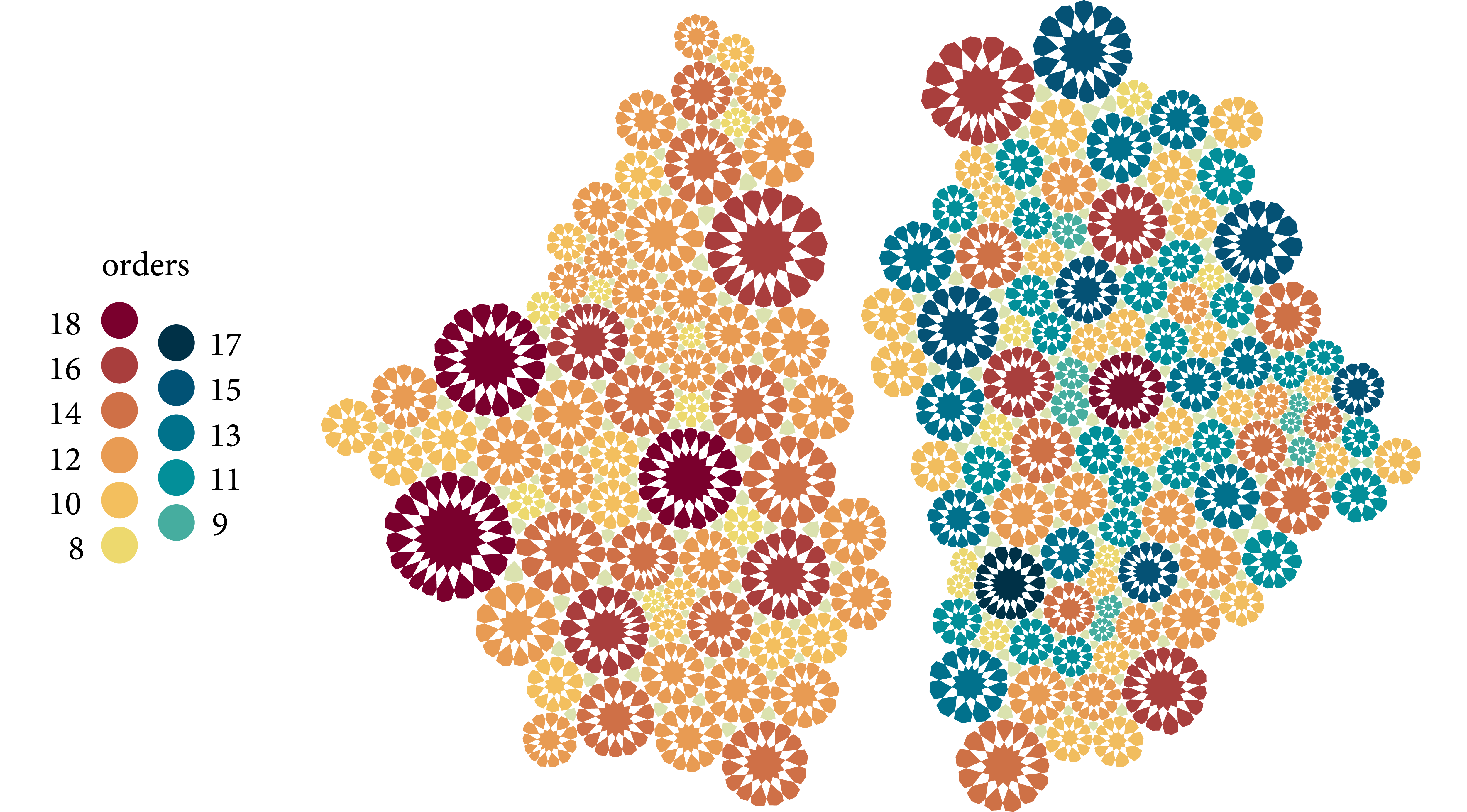}
  \caption{\label{fig:colour-by-size}Generative designs constructed from Delaunay triangulations of random points sets, without (left) and with (right) bowtie gadgets.}
\end{figure}
\section{Results}
\label{sec:results}

\begin{figure}[h]
\centering
  \includegraphics[width=\linewidth]{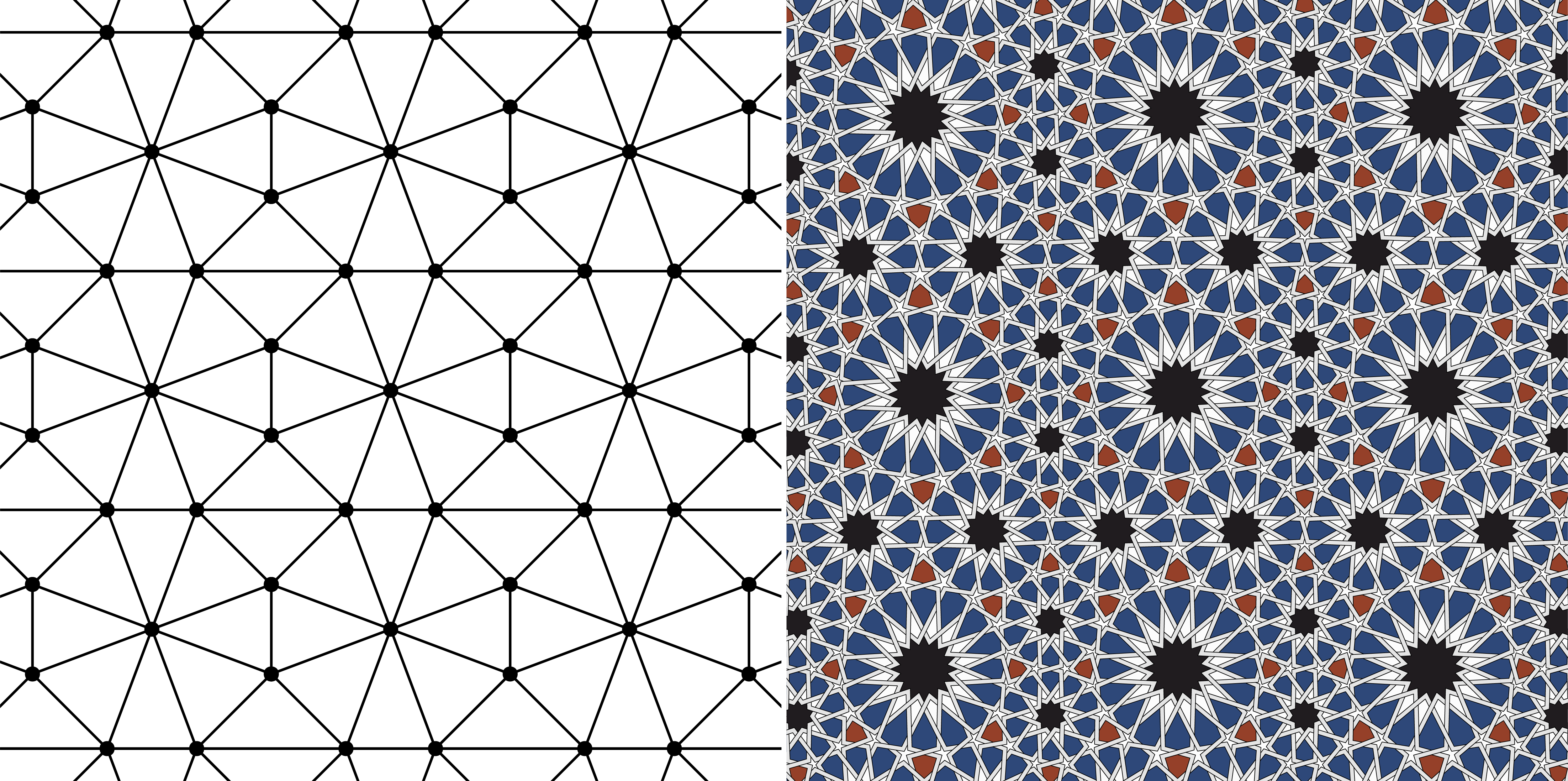}
  \caption{\label{fig:ordered-pattern}A highly structured composition (right) generated from a finite subset 
    of a conceptually periodic complex (left).  Although the circle packing is not 
    constructed in a toroidal domain as in \sect{sec:periodicity}, the resulting
    composition appears close to periodic.}
\end{figure}

We demonstrate the versatility of our technique by presenting a range of freeform designs. For stylized results such as the filled composition in \fig{fig:colour-by-size} and the interlaced design in \fig{fig:ordered-pattern}, we adapt the rendering algorithms described by Kaplan in Bonner's text~\cite[Sec. 4.5]{Bonner2017}.

\begin{figure}[h]
  \includegraphics[width=\linewidth]{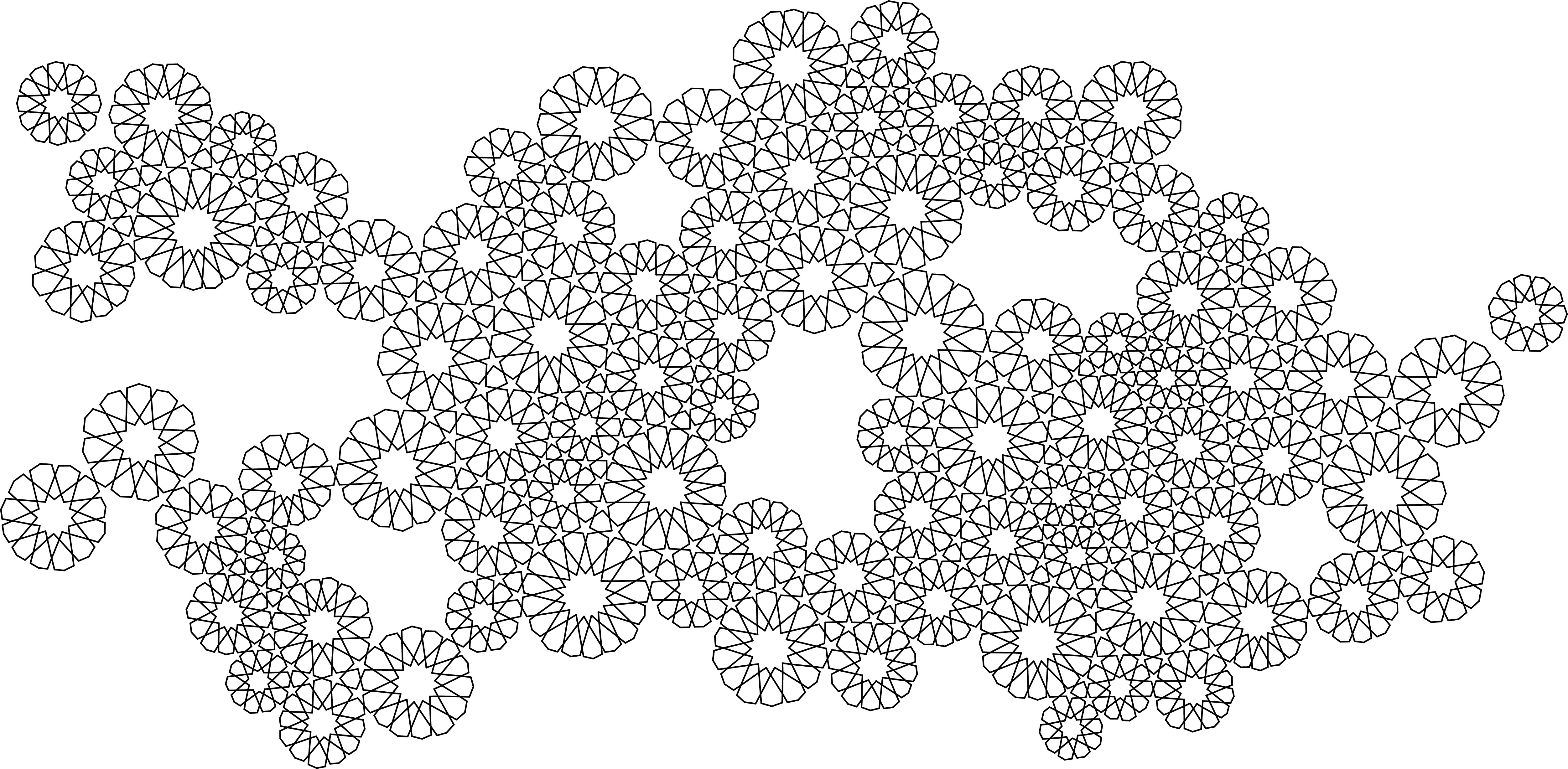}
  \caption{\label{fig:generative}A generative design in which the user has manually chosen to keep a subset of rosettes from an initial arrangement, producing a more dynamic composition
    with an irregular boundary and internal voids.}
\end{figure}

Our method places no constraints on the input complex, giving users considerable control over the output design.
Fully generative designs can be created using Delaunay triangulations of random point sets, leading to arrangements of rosettes with various orders (\fig{fig:colour-by-size}, left). We can further increase the number of possible charges and broaden the expressiveness of our technique by inserting random gadgets (\fig{fig:colour-by-size}, right). Of course, an artist may select a subset of rosettes in a generative design to craft a desired high-level composition (\fig{fig:generative}).

On the other hand, we can begin with a highly structured complex and obtain a repetitive final design 
(Figs.~\ref{fig:freeform-classical} and~\ref{fig:ordered-pattern}), or use a toroidal complex to produce a truly periodic 
pattern (\fig{fig:periodic-pattern-complex}d). In principle, these approaches could be used to produce exactly or approximately periodic drawings of many historical Islamic geometric patterns. However, we have not attempted to catalogue exactly which ones are possible because existing construction
techniques are much better suited to the task of drawing them.

\begin{figure*}[h]
\centering
  \includegraphics[width=\linewidth]{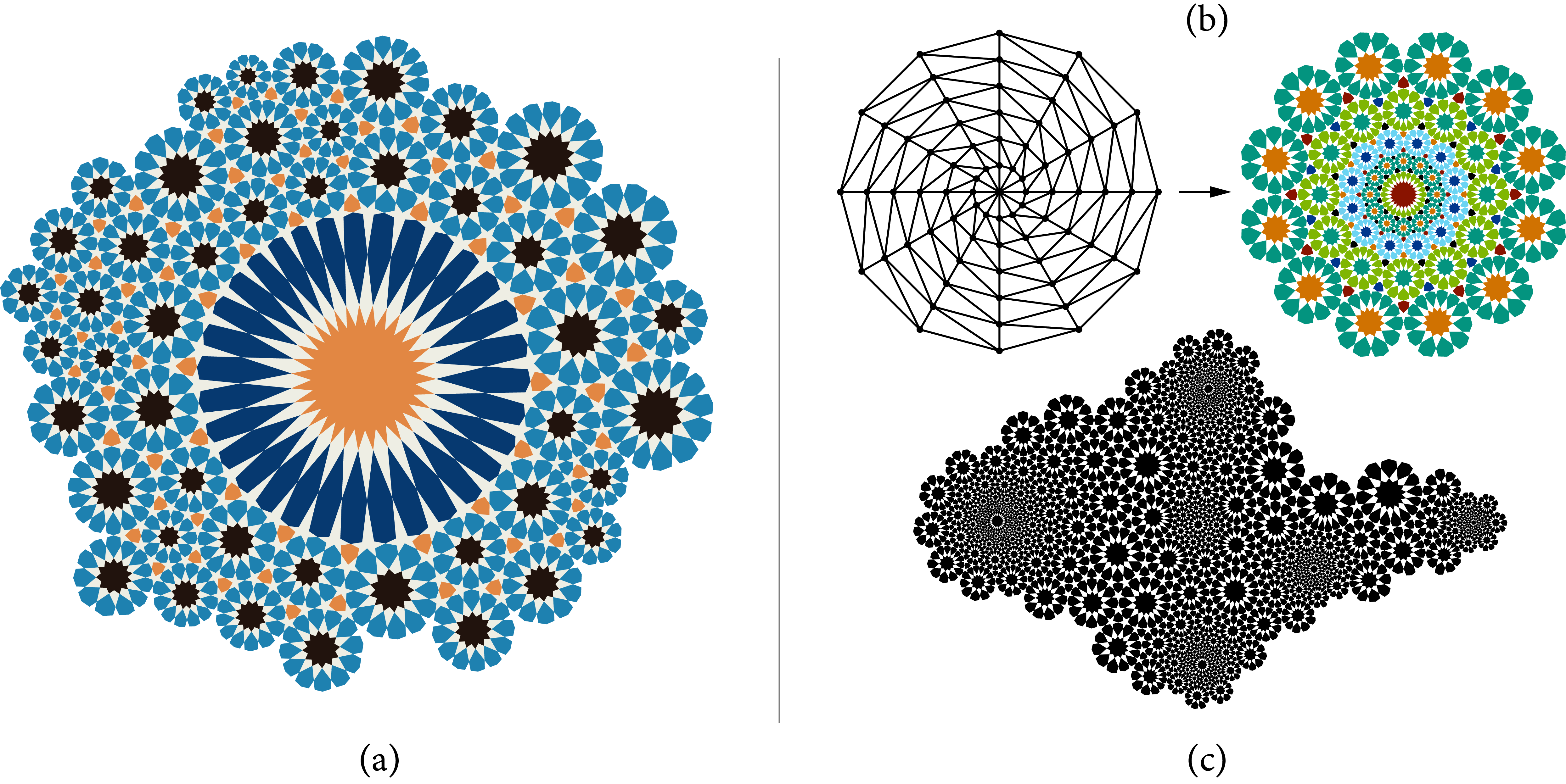}
  \caption{\label{fig:high-order}A tuned design with a high-order rosette (a), for which $\tau = 0.96$ and $\alpha = 0.75$, and a composition (c) incorporating multiple instances of a web-like sub-complex (b).}
\end{figure*}

In between the extremes of full control and generative randomness,
we can insert carefully constructed subgraphs into a complex to create a single high-order rosette
(\fig{fig:high-order}a), or create appealing local arrangements of rosettes (\fig{fig:high-order}b,c).
Another way to balance order and chaos is to place random gadgets within an otherwise ordered grid
(\fig{fig:grid-gadgets}).

The high-order rosette in \fig{fig:high-order}a is a special case, in that it requires hand-tuning.
Recall that adjacent circles with widely varying radii can produce distorted motifs (\fig{fig:circle-selection}).
To produce a satisfactory large rosette, we manually set the $\tau$ and $\alpha$ values for its cyclic
polygon for a better fit with the surrounding geometry. 

Close examination of many of our results reveals small geometric discrepancies of the kind illustrated starkly in \fig{fig:circle-selection}. When rosettes have arms that vary too dramatically in width or length, they disrupt the elegance of a pattern and the feeling of `inevitability' in its construction. There are several places in our work where we choose global constants like
$\tau$ that produce acceptable results in general without adapting to the detailed geometry
of local parts of individual designs. The large rosette in \fig{fig:high-order}a gives one
clear example of where local adjustments can improve a design. In future work, we would like to
explore more fine-grained constructions that can enhance the quality of every rosette based on
the configuration of the circle packing in its immediate neighbourhood.

\section{Conclusion}
\label{sec:conclusion}

We presented a robust method for constructing freeform Islamic geometric patterns comprising rosettes of unusual orders. Our technique relies on the theory of circle packings, giving us a principled geometric scaffolding from which to develop a polygonal patch and then motifs. The user controls the initial complex and any gadgets in it, allowing for significant creative freedom in the design process. Our results are more organic and less repetitive than existing patterns and suggest many ideas for further exploration. We believe they communicate the aesthetic of Islamic geometric patterns while also interpreting them in a non-traditional context. They still manage to convey the `aesthetic delight' that Gombrich discussed~\cite{Gombrich1980}, but with slightly less
boredom and more confusion, paving the way for more artistic applications of these designs. The work enhances our understanding of traditional patterns and reveals new opportunities---freeform
or otherwise---for both ornamentation and art-making. 

\section*{Acknowledgements}
This research was supported by the Natural Sciences and Engineering Research Council of Canada and the Cheriton School of Computer Science at the University of Waterloo.

\bibliographystyle{tfs}
\bibliography{main}

\begin{thebibliography}{10}
\providecommand{\MR}{\relax\unskip\space MR }
\providecommand{\url}[1]{\normalfont{#1}}
\providecommand{\urlprefix}{Available at }

\bibitem{AlAjlouni2012}
R.A. Al~Ajlouni, \emph{The global long-range order of quasi-periodic patterns
  in islamic architecture}, Acta Crystallographica Section A 68 (2012), pp.
  235--243. \urlprefix\url{https://doi.org/10.1107/S010876731104774X}.

\bibitem{Arnheim1972}
R. Arnheim, \emph{Art and Visual Perception: A Psychology of the Creative Eye},
  University of California Press, Berkeley and Los Angeles, California, 1974.

\bibitem{Bonner2017}
J. Bonner, \emph{Islamic Geometric Patterns: Their Historical Development and
  Traditional Methods of Construction}, Springer-Verlag, New York, 2017.

\bibitem{BP2012a}
J. Bonner and M. Pelletier, \emph{A 7-Fold System for Creating Islamic
  Geometric Patterns Part 1:~Historical Antecedents}, in \emph{Proceedings of
  Bridges 2012: Mathematics, Music, Art, Architecture, Culture}, R. Bosch, D.
  McKenna, and R. Sarhangi, eds., Phoenix, Arizona. Tessellations Publishing,
  2012, pp. 141--148. Available at
  \url{http://archive.bridgesmathart.org/2012/bridges2012-141.html}.

\bibitem{Brewer2022}
S.G. Brewer, M. Zha, and S. Neno, \emph{Generating Families of Islamic Star
  Rosette Patterns Based on k-Uniform Tilings}, in \emph{Proceedings of Bridges
  2022: Mathematics, Art, Music, Architecture, Culture}, D. Reimann, D. Norton,
  and E. Torrence, eds., Phoenix, Arizona. Tessellations Publishing, 2022, pp.
  391--394.
  \urlprefix\url{http://archive.bridgesmathart.org/2021/bridges2022-391.html}.

\bibitem{Broug2013}
E. Broug, \emph{Islamic Geometric Design}, Thames \& Hudson, London, 2013.

\bibitem{Castera1999}
J.M. Castera, \emph{Arabesques: Decorative Art in Morocco}, ACR Edition, Paris,
  1999.

\bibitem{Castera2010}
J.M. Castera, \emph{From the Angle of Quasicrystals}, in \emph{Proceedings of
  Bridges 2010: Mathematics, Music, Art, Architecture, Culture}, G.W. Hart and
  R. Sarhangi, eds., Phoenix, Arizona. Tessellations Publishing, 2010, pp.
  215--222. Available at
  \url{http://archive.bridgesmathart.org/2010/bridges2010-215.html}.

\bibitem{Castera2016}
J.M. Castera, \emph{Another look at Pentagonal Persian Patterns}, in
  \emph{Proceedings of Bridges 2016: Mathematics, Music, Art, Architecture,
  Education, Culture}, E. Torrence, B. Torrence, C. S\'equin, D. McKenna, K.
  Fenyvesi, and R. Sarhangi, eds., Phoenix, Arizona. Tessellations Publishing,
  2016, pp. 325--330. Available at
  \url{http://archive.bridgesmathart.org/2016/bridges2016-325.html}.

\bibitem{CollinsStephenson2003}
C.R. Collins and K. Stephenson, \emph{A circle packing algorithm},
  Computational Geometry 25 (2003), pp. 233--256.

\bibitem{Cromwell2021a}
P. Cromwell, \emph{Looking at islamic patterns i: The perception of order},
  PsyArXiv (2021). \urlprefix\url{psyarxiv.com/qhg3f}.

\bibitem{Cromwell2010}
P.R. Cromwell, \emph{Islamic geometric designs from the topkapı scroll i:
  unusual arrangements of stars}, Journal of Mathematics and the Arts 4 (2010),
  pp. 73--85. \urlprefix\url{https://doi.org/10.1080/17513470903311669}.

\bibitem{Cromwell2013}
P.R. Cromwell, \emph{On irregular stars in islamic geometric patterns}, 2013.
  Available at
  \url{https://girih.wordpress.com/on-irregular-stars-in-islamic-geometric-patterns/},
  accessed 12 August 2021.

\bibitem{Cromwell2015}
P.R. Cromwell, \emph{Cognitive bias and claims of quasiperiodicity in
  traditional islamic patterns}, Math Intelligencer 37 (2015), pp. 30--44.

\bibitem{Eppstein}
D. Eppstein, \emph{{CirclePack.py}} (2012).
  \urlprefix\url{https://www.ics.uci.edu/~eppstein/PADS/CirclePack.py}.

\bibitem{Gailiunas2020}
P. Gailiunas, \emph{Near-miss Star Patterns}, in \emph{Proceedings of Bridges
  2020: Mathematics, Art, Music, Architecture, Education, Culture}, C. Yackel,
  R. Bosch, E. Torrence, and K. Fenyvesi, eds., Phoenix, Arizona. Tessellations
  Publishing, 2020, pp. 27--34.
  \urlprefix\url{http://archive.bridgesmathart.org/2020/bridges2020-27.html}.

\bibitem{Gieseke2021}
L. Gieseke, P. Asente, R. M\v{e}ch, and M. Benes Bedrich an d~Fuchs, \emph{A
  survey of control mechanisms for creative pattern generation}, Computer
  Graphics Forum 40 (2021), pp. 585--609.
  \urlprefix\url{https://onlinelibrary.wiley.com/doi/abs/10.1111/cgf.142658}.

\bibitem{Gombrich1980}
E. Gombrich, \emph{The Sense of Order: A study in the psychology of decorative
  art}, second revised ed., Phaidon Press, Oxford, 1994.

\bibitem{Hamekasi2012}
N. Hamekasi and F. Samavati, \emph{Designing Persian Floral Patterns using
  Circle Packing}, in \emph{Proceedings of the International Conference on
  Computer Graphics Theory and Applications and International Conference on
  Information Visualization Theory and Applications (GRAPP/IVAPP)}, P. Richard,
  M. Kraus, R.S. Laramee, and J. Braz, eds., Feb 24-26. SciTePress, 2012, pp.
  135--142. \urlprefix\url{http://dx.doi.org/10.5220/0003850101350142}.

\bibitem{Hankin1925}
E.H. Hankin, \emph{The Drawing of Geometric Patterns in {S}aracenic Art},
  Memoirs of the Archaeological Society of India Vol.~15, Government of India,
  Calcutta, 1925.

\bibitem{Hutcheson1726}
F. Hutcheson, \emph{An inquiry into the original of our ideas of beauty and
  virtue}, Printed by J. Darby in Bartholomew Close, London, 1725,
  \urlprefix\url{https://oll.libertfund.org/title/leidhold-an-inquiry-into-the-original-of-our-ideas-of-beauty-and-virtue-1726-2004\#lf1458\_head\_013}.

\bibitem{Kaplan2009}
C.S. Kaplan, \emph{Semiregular patterns on surfaces}, in \emph{NPAR '09:
  Proceedings of the 7th international symposium on Non-photorealistic
  animation and rendering}, New York. ACM Press, 2009, pp. 35--39.

\bibitem{Kaplan2022}
C.S. Kaplan, \emph{Generative Zellij}, in \emph{Proceedings of Bridges 2022:
  Mathematics, Art, Music, Architecture, Culture}, D. Reimann, D. Norton, and
  E. Torrence, eds., Phoenix, Arizona. Tessellations Publishing, 2022, pp.
  285--288.
  \urlprefix\url{http://archive.bridgesmathart.org/2022/bridges2022-285.html}.

\bibitem{KaplanSalesin2004}
C.S. Kaplan and D.H. Salesin, \emph{Islamic star patterns in absolute
  geometry}, ACM Transactions on Graphics 23 (2004), pp. 97--119.

\bibitem{Lee1987}
A. Lee, \emph{Islamic star patterns}, Muqarnas 4 (1987), pp. 182--197.

\bibitem{LuSteinhardt2007}
P.J. Lu and P.J. Steinhardt, \emph{Decagonal and quasi-crystalline tilings in
  medieval islamic architecture}, Science 315 (2007), pp. 1106--1110.

\bibitem{BP2012b}
M. Pelletier and J. Bonner, \emph{A 7-Fold System for Creating Islamic
  Geometric Patterns Part 2:~Contemporary Expression}, in \emph{Proceedings of
  Bridges 2012: Mathematics, Music, Art, Architecture, Culture}, R. Bosch, D.
  McKenna, and R. Sarhangi, eds., Phoenix, Arizona. Tessellations Publishing,
  2012, pp. 149--156. Available at
  \url{http://archive.bridgesmathart.org/2012/bridges2012-149.html}.

\bibitem{Stephenson2005}
K. Stephenson, \emph{Introduction to circle packing: The theory of discrete
  analytic functions}, Cambridge University Press, New York, 2005.

\bibitem{WR2007}
B. Wichmann and J. Rigby, \emph{A penrose-type islamic interlacing pattern},
  Visual Mathematics 9 (2007). Avaiable at
  \url{http://symmetry-us.com/Journals/wichmann2007/penrose.html}, accessed 12
  August 2021.

\bibitem{ZPA1992}
H. Zabrodsky, S. Peleg, and D. Avnir, \emph{Continuous symmetry measures},
  Journal of the American Chemical Society 114 (1992), pp. 7843--7851.

\end{thebibliography}
\end{document}